\documentclass[a4paper,11pt]{article}
\pdfoutput=1 

\usepackage{jcappub} 

\usepackage[T1]{fontenc} 

\usepackage{graphicx, epsfig,dsfont,subfigure}
\usepackage{multirow} 

\def \A {{\mathbb A}}
\def \K {{\mathbb K}}
\def \R {{\mathbb R}}
\def\case#1/#2{\frac{#1}{#2}}
\def\DerN#1{\frac{d #1}{d N}}

\title{\boldmath A new approach to the analysis of the phase space of $f(R)$-gravity.}

\author[a]{S. Carloni}

\affiliation[a]{Centro Multidisciplinar de Astrofisica - CENTRA,
Instituto Superior Tecnico - IST,
Universidade de Lisboa - UL,
Avenida Rovisco Pais 1, 1049-001, Portugal.}

\emailAdd{sante.carloni@tecnico.ulisboa.pt}

\abstract{We propose a new dynamical system formalism for the analysis of $f(R)$ cosmologies. The new approach eliminates the need for cumbersome inversions to close the dynamical system and allows the analysis of the phase space of $f(R)$-gravity models which cannot be investigated using the standard technique. Differently form previously proposed similar techniques, the new method is constructed in such a way to associate to the fixed points scale factors, which contain four integration constants (i.e. solutions of fourth order differential equations). In this way a new light is shed on the physical meaning of the fixed points. We apply this technique to some $f(R)$ Lagrangians relevant for inflationary and dark energy models. 
}

\begin{document}
 \maketitle

\section{Introduction}
Since the first formulation of General Relativity (GR), many extensions of the original Einstein equations have been investigated. The reasons of the interest in such theories are quite disparate: from the first attempts to unify geometrically the electromagnetic and gravitational interaction started by Weyl \cite{Weyl}, to the understanding of the corrections to the gravitational action typical of quantum field theory in curved spacetime and  fundamental unification schemes \cite{Donoghue:1995cz}, to the attempts to give a complete geometric explanation of the dark phenomenology. 

Among these extensions, the class of theories called $f(R)$-gravity \cite{f(R) review} is the simplest realisation of higher order gravity (order four) and  has an important role as a natural model for inflation \cite{Starobinsky:1980te}. More recently, it was shown to have also an interesting (and very debated) role as geometrical dark energy model \cite{f(R)-DE}. In dealing with these theories, the  necessity to solve fourth order  differential equations is the source of  the difficulties in the true understanding of the features of their cosmology. Such problems were the origin of the development of a series of methods to indirectly analyse the physical properties of $f(R)$ cosmologies. 

Dynamical System Approach (DSA) has proven to be one of the most effective of such methods. DSA has a number of different realisations \cite{Belisnky, Misner}. In the following we shall consider the one proposed in the 1970's by Collins \cite{Collins} and successively developed by Wainwright, Ellis and Uggla. This version of DSA is defined in terms of dimensionless, expansion normalised variables and allows a very clear physical interpretation of the results. DSA has had a key role in analysing Bianchi models \cite{DSA Bianchi}  and  minimally coupled scalar tensor \cite{Copeland:1997et} in the context of GR and has allowed the exploration of the cosmology of a number of modifications of Einstein's theory \cite{Carloni:2004kp,ModGravDSA}. In \cite{Carloni:2004kp}, the model $f(R)= \chi R^n$ was analysed  with this method for the fist time. Later on the method was extended in \cite{Amendola:2006we,Carloni:2007br}  to the case of a generic $f(R)$.  DSA allowed for the first time to analyse in detail the phase space for $f(R)$ cosmologies making some generals statements on these cosmologies and revealing a number of interesting attractor solutions.

In spite of this success the method above has some unsatisfactory aspects. First of all the possibility to analyse a given $f(R)$ theory depends on the exact resolution of an algebraic equation of generic order, or, in the most complicated cases, a transcendental equation. Such operation not only limits the set of possible Lagrangians which can be analysed with DSA, but also introduces a number of singularities, so that in many cases the dynamical system is not of class $C^1$.  

Another important problem is that the choice of the dynamical system variables was made in such a way to obtain in the fixed points solutions with only two integration constants. This implies that these solutions correspond to the general solutions of  the fourth order cosmological model where two integration constants have been set to zero. This is problematic because without knowledge of the full solution in a fixed point it is impossible to characterise the correct behaviour of these cosmologies when these fixed points are nodes.

In addition, since the dynamical system variables are not always independent from each other, the dynamical system might present fixed points which 
can correspond to inconsistent conditions in the cosmological equations i.e. to have $f(R)=0$ and $R=0$ for a function $f$ for which $f(0)\neq0$. The presence of these points must be a spurious effect due to the way in which he DSA is constructed.

In this paper we propose a new DSA to deal with the cosmology of $f(R)$ gravity. This method is built in such a way to avoid the necessity of  solving exactly algebraic/transcendental equations and to give, in the fixed points, solutions of the cosmological equations which contain four integration constants. We will show that the new formulation contains the results of the old method but reveals unsuspected additional features of the evolution of $f(R)$ cosmologies.

The paper is organised as follow. In section 2 we will give the basic equations. In section 3 we will introduce the original DSA for $f(R)$-gravity. Section 4 is dedicated to the construction of the new DSA, and  in Section 5 the new method is applied to some interesting model of $f(R)$ gravity. Section 6 is dedicated to the conclusions.

Unless otherwise specified, natural units ($\hbar=c=k_{B}=8\pi G=1$)
will be used throughout this paper, Latin indices run from 0 to 3.
The symbol $\nabla$ represents the usual covariant derivative and
$\partial$ corresponds to partial differentiation. We use the
$(-,+,+,+)$ signature and the Riemann tensor is defined by
\begin{equation}
R^{a}{}_{bcd}=W^a{}_{bd,c}-W^a{}_{bc,d}+ W^e{}_{bd}W^a{}_{ce}-
W^f{}_{bc}W^a{}_{df}\;,
\end{equation}
where the $W^a{}_{bd}$ are the Christoffel symbols (i.e. symmetric in
the lower indices), defined by
\begin{equation}
W^a_{bd}=\frac{1}{2}g^{ae}
\left(g_{be,d}+g_{ed,b}-g_{bd,e}\right)\;.
\end{equation}
The Ricci tensor is obtained by contracting the {\em first} and the
{\em third} indices
\begin{equation}\label{Ricci}
R_{ab}=g^{cd}R_{acbd}\;.
\end{equation}
Finally the Hilbert--Einstein action in the presence of matter is
given by
\begin{equation}
{\cal A}=\int d x^{4} \sqrt{-g}\left[\frac{1}{2}R+ L_{m}\right]\;.
\end{equation}

\section{Basic Equations}
In this paper we are going to deal only with metric $f(R)$ theories, which are characterized by the Action
\begin{equation}\label{lagr f(R)}
\mathcal{S}=\int d^4 x \sqrt{-g}\left[ f(R, \bar{\alpha}, \bar{\beta}...)+{\cal L}_{m}\right]\;,
\end{equation}
where $\mathcal{L}_m$ represents the matter contribution. In general the function $f$ is considered an analytic function of the Ricci scalar $R$ and contains  a set of  additional dimensional parameters indicated by barred Greek letters.

Varying the action with respect to the metric gives the generalisation of the Einstein equations:
\begin{equation}\label{eq:einst}
 f'G_{ab}=T
_{ab}^{m}+\frac{1}{2}g_{ab} \left(f-R f'\right) +(g_{a}^{~c}g_{b}^{~d}-
g_{ab}g^{cd})\nabla_c\nabla_d f'\;,
\end{equation}
where $G_{ab}$ is the Einstein tensor,  $f=f(R, \bar{\alpha}, \bar{\beta}...)$, $f'= \displaystyle{\frac{d f(R, \bar{\alpha}, \bar{\beta}...)}{dR}}$, and
$\displaystyle{T^{M}_{ab}=\frac{2}{\sqrt{-g}}\frac{\delta
(\sqrt{-g}\mathcal{L}_{m})}{\delta g_{ab}}}$ represents the
stress energy tensor of standard matter. These equations reduce to
the standard Einstein field equations when $f(R, \bar{\alpha}, \bar{\beta}...)=\bar{\alpha} R$ with $\bar{\alpha}=1/2$. 

Our treatment will consider only  homogeneous and isotropic spacetimes i.e. Friedmann Lema\^{\i}tre Robertson Walker (FLRW) metrics:
\begin{equation}\label{frw}
 ds^2 = -dt^2 + a^2(t)\left[ {dr^2 \over 1-kr^2} + r^2 (d\theta^2 +
\sin^2\theta d\phi^2)\right]\;,
\end{equation}
where $a$ is the scale factor and  $k$ the spatial curvature. We also assume that the cosmic fluid is a prefect fluid 
with equation of state $p=w \mu$  with $0\leq w\leq1$. It is common to write the field equations (\ref{eq:einst})
in the metric \eqref{frw} as two equation resembling the Raychaudhuri and the Friedmann equations in GR
\begin{align}\label{CosmEq}
\begin{split}
& H^{2}+\frac{k}{a^2} =
\frac{1}{3f'}\left\{\frac{1}{2}\left[f'R-f\right]-3H\dot{f'}+\mu_{{m}}\right\}\,,\\
&2\dot{H}+H^{2}+\frac{k}{a^2} =
-\frac{1}{f'}\left\{\frac{1}{2}\left[f'R-f\right]+\ddot{f'}-3H\dot{f'}+\,p_{{m}}\right\}\,,
\end{split}
\end{align}
where
\begin{equation}
R\,=\,6\left(\dot{H}+2H^{2}+\frac{k}{a^2}\right)\,,\label{R}
\end{equation}
$H\equiv\dot{a}/a$, the prime represents the derivative with respect to $R$ and the
``dot" is the derivative with respect to $t$. The two equations \eqref{CosmEq} are not independent: the second can be obtained deriving the first with respect to the cosmic time $t$ once the Bianchi identities for
$T^{m}_{\mu\nu}$ is considered. In FLRW these identities take the form
\begin{equation}\label{BianchiMatt}
\dot{\mu}_{m}+3H(\mu_{ m}+p_{m})=0\;,
\end{equation}
which is the same as the GR energy conservation equation for the cosmic fluid.

\section{The Original Dynamical Systems Approach for $f(R)$ gravity (in brief).}
Using the equations above we can formulate the cosmic evolution in terms of dynamical systems \cite{Amendola:2006we,Carloni:2007br} (beware of the differences in signature!). Introducing the general dimensionless variables \,:
\begin{eqnarray}\label{var}
\begin{split}
x = \frac{\dot{f'}}{H f' }, \qquad y = \frac{R}{6 H^2},  \qquad z = \frac{f}{6 H^2 f' }, \\
\Omega = \frac{\mu_m}{3 H^2 f' },  \qquad K =\frac{k}{a^2 H^2}\;,
\end{split}
\end{eqnarray}
where $\mu_m$ represents the energy density of a perfect fluid
that is present in the model. As customary, we also  define the logarithmic (dimensionless) ``time variable'' $N=\ln a$. Note that in choosing this time variable we are assuming that we represent the phase space for $H>0$ i.e. we are considering only expanding cosmologies\footnote{ 
The contracting case can also be considered using the time variable $M=-N$. Since by definition almost all the variables \eqref{var} are invariant under a change of sign of $H$ the phase space for $H<0$ will have a strict resemblance with the one with $H>0$. The $H<0$ part of the phase space can be important to analyse, for example bouncing scenarios. In the following we will not consider this part of the phase space focusing only on expanding cosmologies, leaving this analysis for a future work.}.   In this variables the cosmological equations (\ref{CosmEq}) are equivalent to the
autonomous system:
\begin{eqnarray}\label{DynsysNoK}
\frac{dx}{d N} &=&\,x (\Omega -z-2)-2 x^2+2 (y-2 z)+(1-3 w) \Omega , \nonumber \\
\frac{dy}{d N} &=&y \,[(\Gamma -2) x+2 \Omega -2 z+2],  \\
\frac{dz}{d N} &=&z\,(2-3 x-2 z+2 \Omega) z+x\, y \Upsilon, \nonumber \\
\frac{d\Omega}{d N} &=& \Omega \,\,(2 \Omega-3 x-2 z-1 -3 w), \nonumber \\
0 &= &1+ K + x + z - y - \Omega \;. \nonumber
\end{eqnarray}
The quantity $\Upsilon$ is defined as
\begin{equation}\label{q}
\Upsilon\,\equiv\,\frac{f'}{Rf''}\,.
\end{equation}
Since $\Upsilon$ is a function of $R$ only,  the problem of obtaining $\Upsilon=\Upsilon(x,y,z,\Omega)$ is reduced to the problem of writing $R=R(x,y,z,\Omega)$. This can be achieved from the definitions \eqref{var}:
\begin{equation}\label{Pre-r}
\frac{y}{z}=\frac{Rf'}{f}\,.\,
\end{equation}
Solving the above equation for $R$ allows one to write $R$ in terms of $y$ and $z$ and close the system \eqref{DynsysNoK}. It is clear that the properties of \eqref{Pre-r} determine the possibility of closing (and therefore analysing) the system \eqref{DynsysNoK} as well as some of the properties of this system i.e. the differential structure. One example is $f(R)=R^p\exp(q R)$ for which 
\begin{equation}
\Upsilon= \frac{yz}{y^2-p z^2}.
\end{equation}
In this case it is evident that the system is not $C(1)$ as the curve $y^2-p z^2=0$ is singular. This fact can have serious repercussions on the properties of the flow \cite{Carloni:2007br}.
\section{The new Dynamical Systems Approach.}
We will now start to construct the new approach. Before we define the dynamical system variables we will need, however, two preliminary steps. The first one will concern the form of the action and the second one will be the introduction of new (cosmic) parameters. These (re-)definitions will be the cornerstones of the new method.
\subsection{The form of the action.}\label{NewAction}
In dealing with dynamical system it is crucial to gain an understanding (and control) over the dimensional structure of the theory we are considering. The reason is that the number of dynamical system variables needed for  DSA will also depend on the number of dimensional constants present in the theory. To construct the new method, therefore, we will rewrite the action \eqref{lagr f(R)} in a special form. In particular we will introduce a  constant $R_0$ such that the product $RR_0$ is dimensionless. In addition, we will also introduce some dimensionless parameter in the form of Greek letters which will represent the ratio between the coupling constant of the additional invariants in the theory  and (a power of) $R_0$. In this  way the  \eqref{lagr f(R)} can be written as 
\begin{equation}\label{lagr f(R)R0}
\mathcal{S}=\int d^4 x \sqrt{-g}\left[ f(R R_0, \alpha...)+{\cal L}_{m}\right]\;,
\end{equation}
where $f$ has the same properties of the one in \eqref{lagr f(R)} and $R_0$ will be assumed non-negative. The main reason behind the formulation above is that in this way any $f(R)$ action contains only one dimensional constant. Therefore, instead of defining a dynamical variable for each dimensional constant $(\bar{\alpha}, \bar{\beta},...)$, we only need one dynamical variable related to $R_0$ to analyse the phase space of actions of any complexity. In addition, this setting prevents the appearance of fixed points not consistent with the cosmological equations  which are typical of the original DSA. We will use this formulation \eqref{lagr f(R)R0} of the action as a starting point in the construction of our new DSA.

\subsection{New cosmic parameters.}
Looking for a different way to constrain the properties of the cosmic fluids, Visser  proposed a set of {\it cosmic (or cosmographic) parameters} \cite{Visser,Sahni:2002fz,Alam:2003sc,Dunajski:2008tg}
\begin{equation}\label{VisserPar}
 q=-\frac{\ddot{a}}{a} \;H^{-2}\,,\qquad j=\;\frac{\ddot{a}}{a} \;H^{-3}\,,\qquad s=\;\frac{\ddddot{a}}{a}\; H^{-4}\,,
\end{equation}
with which one is able to characterize completely  a cosmological model. These quantities are directly related with the Taylor development of the scale factor and this property determines also our capability to measure them. With few exceptions \cite{Sahni:2002fz}, in the case of GR  only the lowest order cosmic parameter have been so far fully exploited, but in the case of $f(R)$-gravity the situation is different: higher order parameters become crucial  and can be used to characterize the evolution of  many important cosmological phenomena e.g. structure formation \cite{Ananda:2008tx}.  

However,  after a quick look to the  $f(R)$ cosmological equations written in terms of these quantities (e.g.  \cite{Capozziello:2008qc,Ananda:2008tx}), one soon realizes that this type of cosmographic parameters are not always the ideal objects to work with. The same happens when one tries to use them to formulate a dynamical system approach: although in principle  the \eqref{VisserPar} are the ideal objects to construct the DSA they do not constitute always an advantageous  set of variables.  

For this reason, it is necessary to look for new sets of  parameters which share the structure of the original cosmographic parameters \eqref{VisserPar} but, at the same time, are more suitable to deal with our specific problem. One possibility is to use the Hubble parameter to define the variables:
\begin{equation} \label{HubbleVarRec}
 \bar{q} =\frac{\dot{H}}{H^2},\qquad \bar{\jmath} =\frac{\ddot{H}H}{\dot{H}^2}, \qquad \bar{s} =\frac{\dddot{H}H^2}{\dot{H}^3}.
\end{equation}
The variables above  have been used in \cite{Carloni:2010ph}  to propose new ways to perform the reconstruction of exact cosmological solutions. Unfortunately also the \eqref{HubbleVarRec} do not prove useful to implement a more powerful dynamical system approach.  Another possibility, which will be adopted in the following,  is to define
\begin{equation} \label{HubbleVarDS}
{\mathfrak q} =\frac{\dot{H}}{H^2},\quad {\mathfrak j} =\frac{\ddot{H}}{ H^2}-\frac{\dot{H}^2}{
   H^3},\quad{\mathfrak s}=\frac{\dddot{H}}{ H^4}+3\frac{\dot{H}^3}{ H^6}-4\frac{ \dot{H}
   \ddot{H}}{ H^5}\,.
\end{equation}
This choice, apparently cumbersome, appears much simpler in terms of the logarithmic time $N$:
\begin{equation} \label{HubbleVarDSN}
{\mathfrak q} =\frac{H_{,N}}{H},\quad {\mathfrak j} =\frac{H_{,NN}}{H},\quad{\mathfrak s}=\frac{H_{,NNN}}{H}\,.
\end{equation}
The (\ref{HubbleVarDSN}) will  be the second cornerstone of the mode we intend to propose.

In terms of ${\mathfrak q}, {\mathfrak j}, {\mathfrak s}$ the Ricci scalar and its derivatives read
\begin{eqnarray}
&& R=6 \left[\left( \mathfrak{q}+ 2\right)H^2 +\frac{k}{a^2} \right],\label{R-Par}\\
&&  \dot{R}= 6 H
   \left\{\left[ \mathfrak{j}+ \mathfrak{q} ( \mathfrak{q}+4)\right] H^2-\frac{2
   k}{a^2}\right\},\\
&&   \ddot{R}=6H^2 \left\{\left[
   \mathfrak{s} +4
   \mathfrak{j} ( \mathfrak{q}+1)+ ( \mathfrak{q}+8) \mathfrak{q}^2\right]H^2+2 (2- \mathfrak{q})\frac{k}{a^2}\right\} ,
\end{eqnarray}
and the cosmological equations \eqref{CosmEq} can be written as
\begin{align}\label{CosmEqVar}
\begin{split}
& H^2(1+\mathfrak{q})+\frac{\mu}{3  f'}+12\frac{k}{a^2}\frac{H^2  f''}{f'}-\frac{f}{6  f'}-\frac{6 H^4
   f'' }{f'} \left[\mathfrak{j}+ \mathfrak{q}(\mathfrak{q}+4)\right]=0,\\
& H^2 + \frac{k}{a^2}-\frac{\mu }{6 f'} (1+3 w)-\frac{f}{6  f'}+\\
&~~~~~+\frac{H^2  f''}{f'} \left\{3 H^2 [ \mathfrak{s}+\mathfrak{j} (4 \mathfrak{q}+5)+  \mathfrak{q}^2 (\mathfrak{q}+9)+4\mathfrak{q}]+6 (1- \mathfrak{q})\frac{k}{a^2}\right\}+\\
&~~~~~+\frac{36 H^2  f^{(3)}}{f'}\left[-\frac{6 k H^2  ( \mathfrak{j} +   \mathfrak{q}^2+4 \mathfrak{q})}{a^2}+\frac{9} {2}H^4 ( \mathfrak{j} +   \mathfrak{q}^2+4 \mathfrak{q})^2-\frac{2 k^2}{a^4}\right],
\end{split}
\end{align}
respectively. Note that now the cosmological equations are equations for $\mathfrak{j}$ and $\mathfrak{s}$ instead of $H$ and $\dot{H}$. In fact, in \eqref{CosmEqVar}, this last quantity  has been substituted according to \eqref{HubbleVarDS}. 

In the system \eqref{CosmEqVar} the Ricci scalar only remains present  in the function $f$.  This means that the equations above should be supplied with an additional constraint given by the \eqref{R-Par}.  This relation will allow to simplify considerably the final dynamical system.

\subsection{The General Method}\label{GenMethod}
The first step to obtain and autonomous system of first order differential equations corresponding to the  \eqref{CosmEqVar} is the definition of the dynamical variables.  In the case of single fluid cosmology\footnote{The approach can be trivially generalised to the multi-fluid case by adding a suitable number of $\Omega$ variables each corresponding to the energy density of the given fluids. Such generalization does not add anything to the understanding of the method and it will not be pursued here.}, we choose the set of variables:
\begin{align} \label{DynVar}
\begin{split}
\mathbb{R}=\frac{R}{6 H^2},\quad \mathbb{K}=\frac{k}{a^2 H^2},\quad \Omega =\frac{\mu }{3H^2  f' },\\ \mathbb{J}=\frac{\mathfrak j}{4},\quad \mathbb{Q}=\frac{3 }{2}{\mathfrak q},\quad \mathbb{A}=R_0H^2\,.
\end{split}
\end{align}
Note that the variable associated to matter does not coincide exactly with the matter density parameter. This is a manifestation of the non-minimal coupling of matter and gravitation typical of $f(R)$-gravity. Also, because of our definition of $R_0$, the variable $\mathbb{A}$ will be always non-negative. In addition to the above variables, we will introduce, like in the original DSA, the logarithmic time variable $N$.
 
With this choice the cosmological equations  \eqref{CosmEqVar} are completely equivalent to the autonomous system:
\begin{align}\label{DynSys}
\begin{split}
&\DerN{\mathbb{R}}= \frac{4}{9} \mathbb{Q} (\mathbb{Q}-3 \mathbb{R}+9)-2 \mathbb{R}+4 \mathbb{J}+4,\\ 
&\DerN{\Omega} =\frac{ \Omega}{18}  \{[9 \mathbb{R}-2 (\mathbb{Q}^2+9\mathbb{Q}+9 \mathbb{J}+9)] {\bf Y}-6 [4\mathbb{Q}+9 (w+1)]\},\\ 
&\DerN{\mathbb J}=\frac{ {\bf Z}}{54 {\bf Y}}\{9 \mathbb{R}-2 [\mathbb{Q} (\mathbb{Q}+9)+9( \mathbb{J}+1)]\}^2+\\
&~~~~~~~~+\frac{ 1}{3 {\bf Y}} [6 {\bf X}+4 \mathbb{Q}-6 \mathbb{R}+3 \Omega(1+3w) +6]+\\ 
&~~~~~~~~+ \frac{ 1}{54}\{9 (3-2 \mathbb{Q})\mathbb{R}+2 (3+2 \mathbb{Q}) [\mathbb{Q} (\mathbb{Q}+15)+9(5 \mathbb{J}-1)]\},\\
&\DerN{\mathbb{Q}}=6\mathbb{J}-\frac{2 \mathbb{Q}^2}{3},\\
&\DerN{\mathbb{K}}=-\mathbb{K}\left(\frac{4 \mathbb{Q}}{3}+2\right),\\
&\DerN{\mathbb{A}}=\frac{4}{3} \mathbb{A} \mathbb{Q}\,,
\end{split}
\end{align}
together with the two constraints
\begin{eqnarray}
&& 1=\Omega-\mathbb{K}+\mathbb{R}-{\bf X}-\left[\left(1+\frac{\mathbb{Q}}{9}\right)\mathbb{Q}-\frac{\mathbb{R}}{2}+\mathbb{J}+1\right]
   {\bf Y},  \label{FriedConstr} \\
&&  \mathbb{R}=\mathbb{K}+\frac{2}{3}\mathbb{Q}+2.\label{Ricci Constr}
\end{eqnarray}
The first corresponds to the Hamiltonian (Friedmann) constraint, which guarantees the conservation of matter energy, and the second is simply the definition of the Ricci scalar. The functions ${\bf X}={\bf X}\left(\mathbb{A},\mathbb{R}\right)$, ${\bf Y}={\bf Y}\left(\mathbb{A},\mathbb{R}\right)$, $ {\bf Z}={\bf Z}\left(\mathbb{A},\mathbb{R}\right)$ are defined respectively as 
\begin{align} \label{XYT}
\begin{split}& {\bf X}\left(\mathbb{A},\mathbb{R}\right)= \frac{f\left(\mathbb{R}, \mathbb{A},\alpha,...\right)}{6 H^2  f'\left(\mathbb{R}, \mathbb{A},\alpha,...\right)},\qquad  {\bf Y}\left(\mathbb{A},\mathbb{R} \right)= \frac{24H^2  f''\left(\mathbb{R}, \mathbb{A},\alpha,...\right)}{  f'\left(\mathbb{R}, \mathbb{A},\alpha,...\right)},\\
& ~~~~~~~~~~~~~~~~~~~~~~ {\bf Z}\left(\mathbb{A},\mathbb{R}\right)= \frac{96 H^4  f'''\left(\mathbb{R}, \mathbb{A},\alpha,...\right)}{f'\left(\mathbb{R}, \mathbb{A},\alpha,...\right)}\,.
\end{split}
\end{align}
These quantities represent the part of the system which depends on the form of the Lagrangian \eqref{lagr f(R)R0}. Note that, differently from the approaches presented in \cite{Amendola:2006we,Carloni:2007br}, in this version of the Dynamical Systems Approach no resolution of algebraic equation is required to close  the system. This means that with the technique proposed here we can in principle analyze {\it all}  $f(R)$ models. 

The two constraints allow us to  eliminate two variables ($\mathbb J$ and $\mathbb Q$) and reduce the total system to: 
\begin{align}\label{DynSysRed}
\begin{split}&\DerN{\mathbb{R}}=2 \mathbb{R} (\mathbb{K}- \mathbb{R}+2)-\frac{4}{{\bf Y}} ({\bf X}+\K-\mathbb{R}-\Omega+1),\\
&\DerN{\Omega}=\Omega  (2-3w+{\bf X}+3 \K-3 \mathbb{R}-\Omega ),\\
&\DerN{\mathbb{K}}=2 \mathbb{K} (\mathbb{K}-\mathbb{R}+1),\\ 
&\DerN{\mathbb{A}}=-2\mathbb{A} (2+\mathbb{K}-\mathbb{R})\,.
\end{split}
\end{align}
It is clear that this system posses a minimum of three invariant submanifolds: (i) $\Omega=0$, (ii) ${\mathbb K} =0$ and (iii) ${\mathbb A} =0$. Depending on the form of ${\bf Y}$ one can also have  a  fourth invariant submanifold in ${\mathbb R} =0$. These invariant submanifolds can be imagined as ``parts" of the phase space which have the property that any orbit that starts in them is trapped. They can have a very specific physical meaning. For example, the existence of the invariant submanifold $\mathbb{K}=0$ tells us that if we start in an orbit with zero spatial curvature we cannot evolve towards a positive or negative curvature parts of the phase space and that the existence of $\Omega=0$ implies that a vacuum cosmology remains vacuum (i.e. standard matter cannot be created or destroyed). The submanifold $\mathbb{R}=0$ represents, instead, the case in which the Ricci scalar is identically zero. Since the equation for $\mathbb{R}$ contains terms with ${\bf Y}$ at denominator, this submanifold can be singular. Finally, the submanifold $\mathbb{A}=0$  corresponds to the case in which the constant $R_0$ is zero and it is of more difficult physical interpretation. It can be thought to correspond to the case in which the gravitational part of the action is identically zero i.e. there is no gravitational interaction \footnote{At this point one might think that reformulating the dynamical system approach distinguishing the dimensional constant in front of the Hilbert-Einstein term and the one(s) of the higher order invariant(s) might ease the interpretation of this kind of invariant submanifolds. Indeed this can be done, but it does not add anything to the phase space analysis. For this reason we will rather keep using the approach presented above, which is more compact.}. Since  $\mathbb{A}$ appears also in ${\bf Y}$,  the submanifold $\mathbb{A}=0$ can also be singular.

The  singular character of the  invariant submanifolds $\mathbb{R}=0$ and $\mathbb{A}=0$, however, does not imply the absence of fixed points.  The existence of fixed points that belong to a singular manifold is one of the exotic properties of dynamical systems which are not of order C(1). The coordinates of  (and therefore the solution associated with) these points can be explained in terms of the limiting form of $f$ for $R,R_0\rightarrow0$. In fact, close to the fixed points in the $\mathbb{R}=0$ and $\mathbb{A}=0$ manifolds the phase space of a given theory $f$ and the one of its limit for $R, R_0\rightarrow0$ can be considered isomorphic and will admit the same fixed points. For these fixed points the stability analysis will not necessarily be the standard one. For example, some of the eigenvalues of these points might diverge. We will impose the condition that actual fixed points will exist if the equations at least of class $C(2)$ i.e. that the dynamical system equations, the cosmological equation and the eigenvalues will be finite for a given fixed point. 

With the above setting, we are ready to analyse the phase space for a given form of $f(R)$\footnote{We should stress here that some general conclusions on the system \eqref{DynSysRed}  could be drawn without specifying the form of $f$. However this can be dangerous. Consider for example a case in which the function ${\bf Y}$ is proportional to $\mathds{R}^{-1}$. In this case the system would admit a fixed point $\mathds{R}=0,\Omega=0,\mathds{K}=0,\mathds{A}=0$ which would be not necessarily present in other cases. For this reason we will refrain form such speculations and will work only on a given forma on $f$.} (and  therefore a form of the functions ${\bf X}$, ${\bf Y}$, ${\bf Z}$). The analysis can be done using the classical dynamical system theory i.e. obtaining the fixed points imposing that the $N$-derivatives of the dynamical variables are zero and using the Hartman-Grobman theorem to determine their stability\footnote{It is clear that, as some free parameters enter in the dynamical equations there will be values of these parameters for which the eigenvalues of the fixed points are zero. Such phenomena are called bifurcations (see e.g. \cite{HirshSmale}) and we will not explore them here. }. Naturally, since the phase space is not compact one will need to perform an asymptotic analysis in order to obtain a full description of the dynamics. Such task, however, will be left for a future work.

The solutions associated to the fixed points can be derived writing the modified Raychaudhuri equation in a fixed point,
\begin{align}\label{RAy3Ord}
\begin{split}
& \mathfrak{s}=\frac{1}{H}\frac{ d^3{H}}{d N^3}= \frac{4}{27} \mathbb{Q}_* \left[\left(2 \mathbb{Q}_*+33\right) \mathbb{Q}_*-9
   \mathbb{R}_*+27\right]+\\ 
& ~~~~~~~~~~-\frac{4}{3} \mathbb{J}_* \left\{\frac{2 {\bf Z}_*}{{\bf Y}_*}[2(\mathbb{Q}_*+9)\mathbb{Q} _*+9(2- \mathbb{R}_*)]+8 \mathbb{Q}_*+15\right\}+\\
&~~~~~~~~~~-\frac{2{\bf Z}_*}{27 {\bf Y}_*}[2(\mathbb{Q}_*+9)\mathbb{Q} _*+9(2- \mathbb{R}_*)]^2 +\frac{8}{ {\bf Y}_*} [2 \mathbb{Q}_*-3 (\mathbb{R}_*+1)]+\\
& ~~~~~~~~~~ -\frac{8{\bf X}_*}{ {\bf Y}_*}-\frac{24 \mathbb{J}_*^2 {\bf Z}_*}{ {\bf Y}_*}+\frac{4  \Omega_*}{{\bf Y}_*}(1+3 w)+2(2-\mathbb{R}_*),
\end{split}
\end{align}
where the asterisk indicates the value of a variable in a fixed point\footnote{It is important to stress here that the \eqref{RAy3Ord} represents the form that the Raychaudhuri equation takes arbitrary close to a fixed point. The meaning of the solutions associated to the fixed points can only be understood correctly in this way. The representation of these solutions that will be given in the following chapter has the only purpose of clarify the nature of the solution of such approximated equations.}. In spite of being a third order differential equation the \eqref{RAy3Ord} is always solvable exactly for $H$. In addition, since \eqref{RAy3Ord} is  a fourth order equation in $a$, the solution associated to the fixed points solution contains four integration constants (and up to three different ``regimes''). As we will see, in the case in which a theory can be treated with both the new and the original method, there is correspondence between the fixed points. This suggests that the original method implicitly sets to zero some integration constants.  We will discover that this can  hide information on the actual meaning of the fixed point, particularly when it is a sink or a source. The other key cosmological quantities can be deduced by the cosmological equations once the \eqref{RAy3Ord} is solved. Specifically one has, from \eqref{BianchiMatt},
\begin{equation}
\mu=a^{-3(1+w)}.
\end{equation}
In literature one often defines the barotropic factor of the high order corrections
considered as an effective fluid.
 
Unfortunately, it will not always  be possible to integrate the \eqref{RAy3Ord} exactly up to an expression for the scale factor. In this case we will rely on considerations on the structure of the equation for $a$  and  numerical integrations to obtain the behaviour of the scale factor and the other key quantities. For this reason we will limit ourselves to give the solutions for the scale factor in terms of plots. 

\section{Examples}
In this section we are going to apply the method described above to some specific models. We will start with two simple ones ($f(R)=R^{n}$, $f(R)=R+\alpha R^{n}$) that are treatable also with the original DSA highlighting common features and differences between these two methods. After that, we will explore other two models (the Starobinsky and the Hu-Sawicki models) that are more physically interesting, but cannot be analysed with the original DSA. 

\subsection{The case of  $R^{n}$-gravity}
As a first check for our new formalism let us consider the first model that has been analysed with the original DSA \cite{Carloni:2004kp}. This model, which is also called sometimes ``$R^{n}$-gravity'', is characterized by an action in which the Ricci scalar appears as a generic power rather than linearly and it constitutes the simplest fourth order modification of GR. 

Putting the action in the form of Section \ref{NewAction} we have
\begin{equation}\label{A R^n}
\mathcal{A}=\int d^4 x \sqrt{-g}\left[ R_0^n R^{n}+{\cal L}_{m}\right]\;.
\end{equation}
Using the original DSA  it was realised that for specific values of the parameter $n$ there could be orbits which naturally present a transition between decelerated and accelerated expansion \cite{Carloni:2004kp}. This model was subsequently subjected to more detailed studies which involved cosmological perturbations \cite{Carloni:2007yv,Ananda:2007xh,Carloni:2010tv,Ananda:2008tx} as well as astrophysical and cosmological tests (e.g. \cite{Capozziello:2003gx,Capozziello:2006ph,Clifton:2005aj}). The result of these investigations and other physical considerations points to the fact that $R^{n}$-gravity is inconsistent with multi-scale observations and should be considered only a toy model.

It is known that in terms of the original DSA, the case $f\propto R^{n}$  is degenerate: two dynamical system variables ($y$ and $z$) coincide. The special character of  $R^{n}$-gravity  is present also in the new approach: since the term  $R_0^n$ appear as a factor of $R^n$ the equation for $\mathbb{A}$ is decoupled and can be excluded. Substituting in the remaining equations the expression  for ${\bf X}$, $\bf Y$ and $\bf Z$
\begin{align}
\begin{split}&{\bf X}= \frac{\R}{ n},\\
&{\bf Y}=\frac{ 4(n-1)}{\R},\\
&{\bf Z}=\frac{8 (n-2) (n-1)}{3\R^2},
\end{split}
\end{align}
we obtain
 \begin{align}\label{DynSysR^n}
 \begin{split}
&\DerN{\mathbb{R}}=  \mathbb{R} \left\{2(2+\K-\mathbb{R})-\frac{1}{n-1}\left[\K+\left(\frac{1}{n}-1\right) \mathbb{R}-\Omega +1\right]\right\},\\
&\DerN{\Omega}=\Omega  \left[3\mathbb{K}+\left(\frac{1}{n}-3\right) \mathbb{R}-\Omega +2-3 w\right]\,,\\
&\DerN{\K}=2 \K(\K+\mathbb{R}+1).
   \end{split}
\end{align}
The system presents  in general three invariant submanifolds: ${\mathbb K} =0$, $\Omega=0$ and ${\mathbb R} =0$. Table  \ref{TavolaRn} contains the standard fixed points of \eqref{DynSysR^n}, together with their associated solution. The stability of the fixed points is shown in the case $w=0$ in Table \ref{TavolaRnStab}.

\begin{figure}[htbp]
\begin{center}
\includegraphics[scale=0.8]{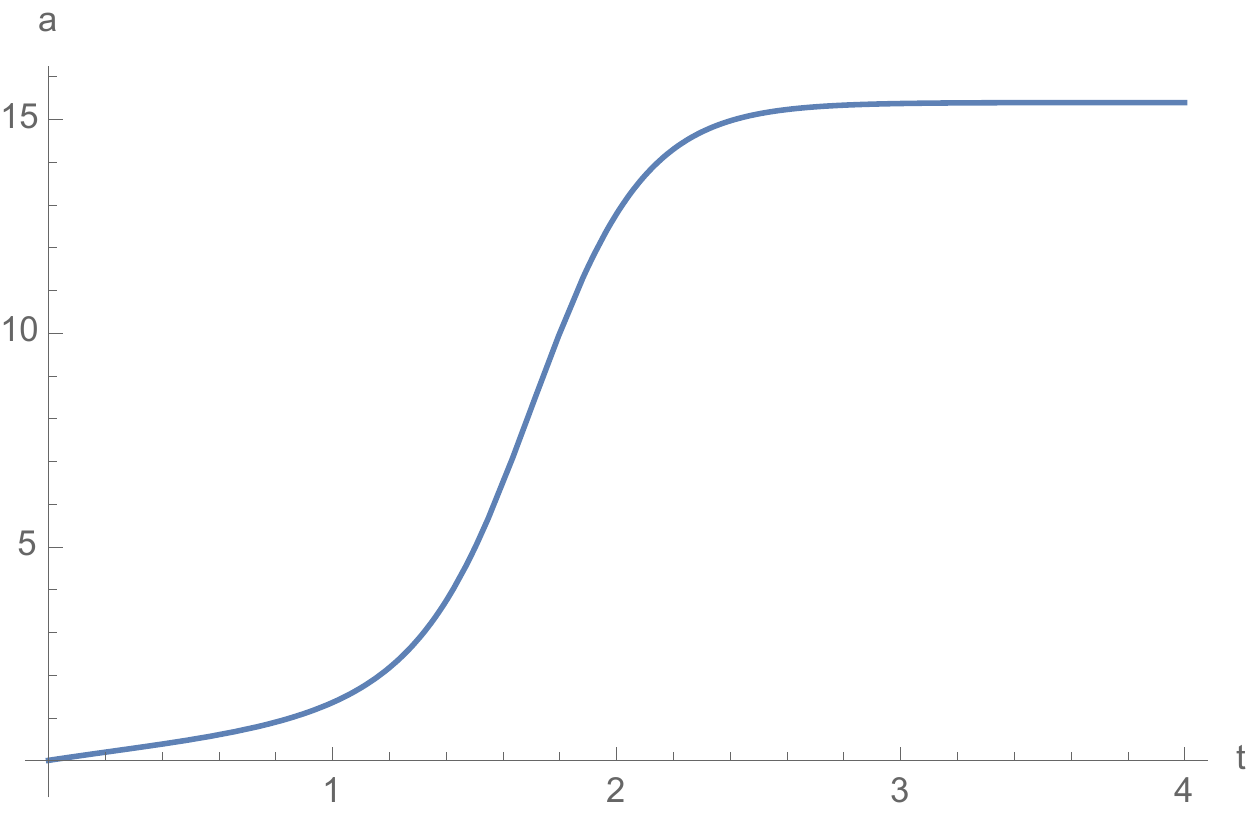}
\caption{Numerical solution of equation \eqref{SolABC}. The constants $H_i$ have all been chosen to be one and the initial condition is $a(0)=0.01$.}
\label{PlotSolABC}
\end{center}
\end{figure}

The solutions associated to the fixed points deserve further discussion. Since now we are solving the full \eqref {RAy3Ord} these solutions will be  specified by a linear differential equation in $a$ which is more complex than the one of the original DSA. For example for the points ${\mathcal A}$, ${\mathcal B}$ and  ${\mathcal C}$ we have\begin{equation}\label{SolABC}
\frac{\dot{a}}{a}=\frac{H_1}{a}+a^{1/2} \left[H_2 \sin \left(\frac{1}{2}\sqrt{3} \log a\right)+H_3 \cos\left(\frac{1}{2}\sqrt{3} \log a\right)\right].
\end{equation}
This equation can be solved exactly (but almost always implicitly) only in the case in which two of the constants $H_i$ are zero. For $H_2$ and $H_3$ zero one has, for example, 
\begin{equation}\label{a=t}
a=a_0 (t-t_0),
\end{equation}
in the other cases the solution can only be expressed in terms of inverses of hypergeometric functions. Looking at the nature of equation \eqref{SolABC} it is clear that for small $a$ the power law behaviour is the dominant component of the solution whereas for large $a$ the hypergeometric behaviour is dominant. A numerical integration of the equation \eqref{SolABC} is given in Figure \ref{PlotSolABC}. 
\begin{figure}[htbp]
\begin{center}
\includegraphics[scale=0.8]{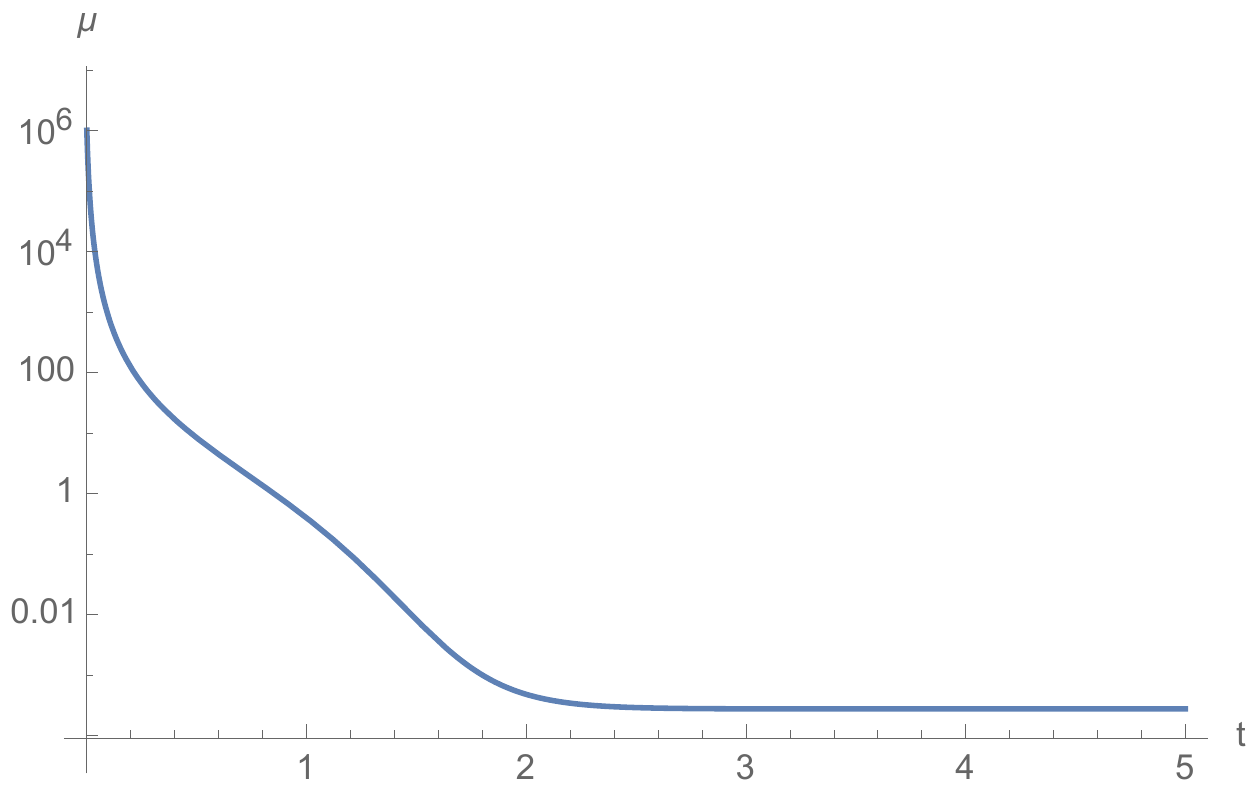}
\caption{Numerical solution for the energy density of \eqref{SolABC}. The constants $H_i$ have all been chosen to be one and the initial condition is $a(0)=0.01$.}
\label{PlotSolABC_mu}
\end{center}
\end{figure}
It is clear that  the solution has sigmoid behaviour: after a power law growth of the type \eqref{a=t}, the expansion rate starts, at first, to increase and then to decrease, to approach eventually  a constant. This results is confirmed by the numerical check of the first derivative of the solution.

The solutions associated to the points ${\mathcal D}$ and  ${\mathcal E}$ are given by the equation
\begin{equation}\label{SolDE}
\frac{\dot{a}}{a}=\frac{H_1}{a^2}+a \left[H_2  \sin \left(\sqrt{3} \log a\right)+H_3 \cos\left(\sqrt{3} \log a\right)\right].
\end{equation}
As before, this equation can be solved exactly only in the case in which two of the constants $H_i$ are zero. For $H_2$ and $H_3$ zero one has, for example, 
\begin{equation}\label{a=t^1/2}
a=a_0 (t-t_0)^{\frac{1}{2}},
\end{equation}
which is dominant only for  small $a$.
\begin{figure}[htbp]
\begin{center}
\includegraphics[scale=0.8]{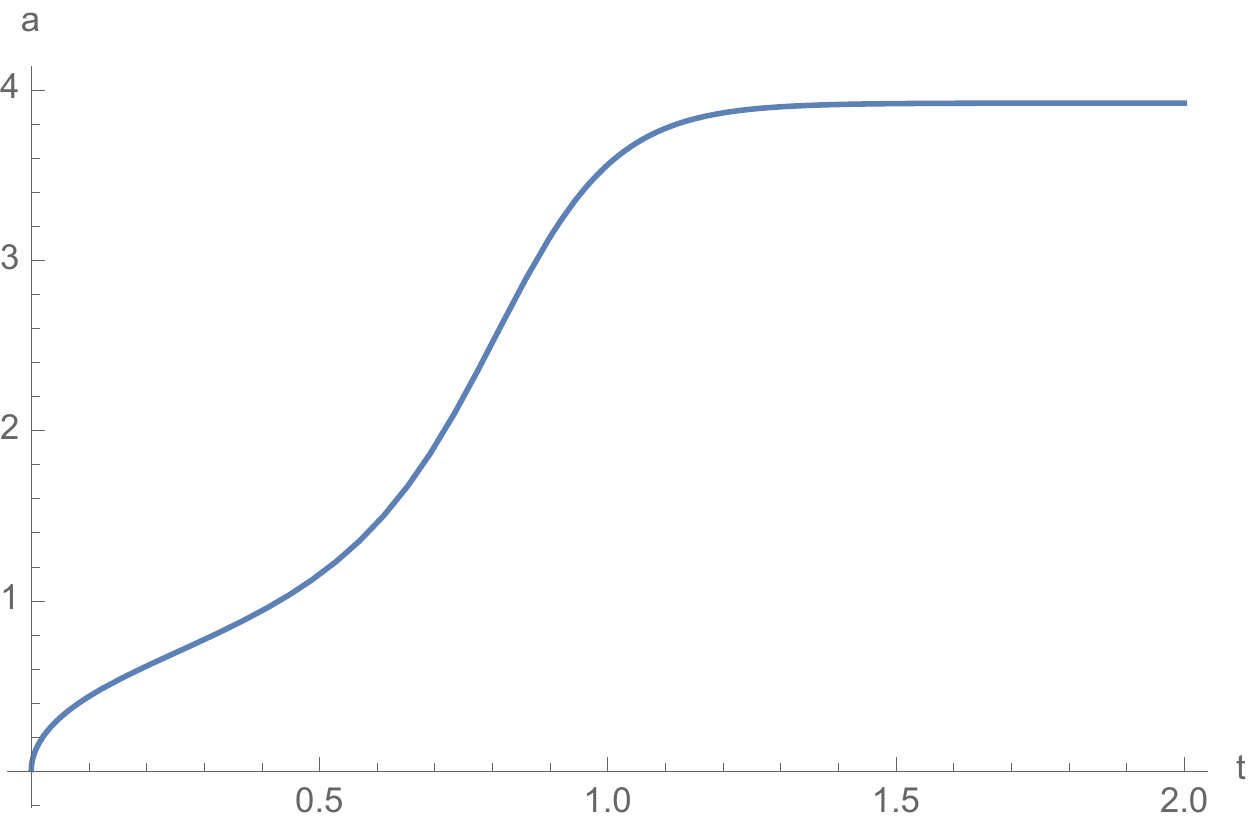}
\caption{Numerical solution of equation \eqref{SolDE}. The constants $H_i$ have all been chosen to be one and the initial condition is $a(0)=0.01$.}
\label{PlotSolDE}
\end{center}
\end{figure}
\begin{figure}[htbp]
\begin{center}
\includegraphics[scale=0.8]{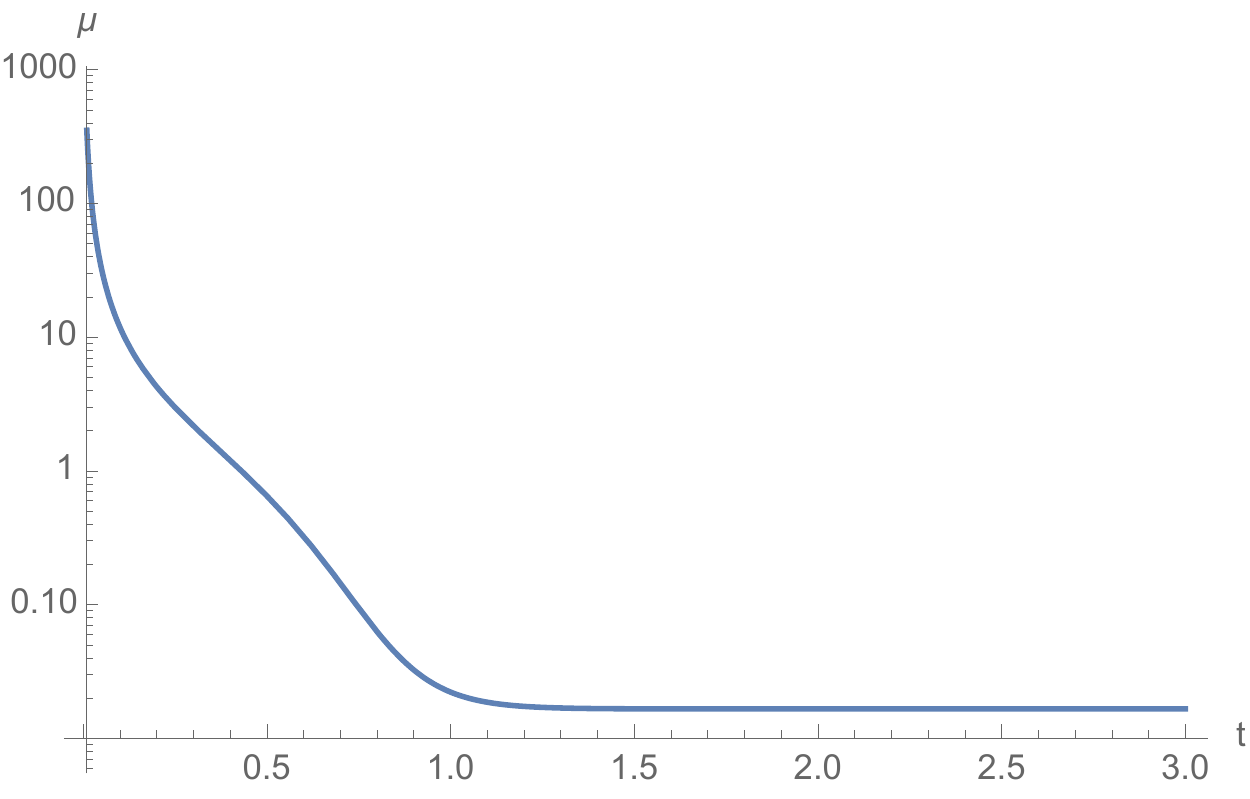}
\caption{Numerical solution for the energy density of \eqref{SolDE}. The constants $H_i$ have all been chosen to be one and the initial condition is $a(0)=0.01$.}
\label{PlotSolDE_mu}
\end{center}
\end{figure}
The numerical integration (Figure \ref{PlotSolDE}) shows that also in this case the solution approaches a constant a late time in a manner similar to the previous one. Both these solutions present two inflection points i.e. changes in the sign of the expansion rate of the universe.  

For ${\mathcal F}$ and  ${\mathcal G}$  instead one has, respectively,
\begin{equation}
\frac{\dot{a}}{a}=\left(H_1+H_2+H_3\right) a^{\frac{n-2}{2 n^2-3 n+1}},
\end{equation}
and 
\begin{equation}
\frac{\dot{a}}{a}=\left(H_1+H_2+H_3\right) a^{-\frac{3 (w+1)}{2 n}},
\end{equation}
which can be integrated exactly to give a pure power law behaviour.

\begin{table}[h]
\begin{center}
\caption{Fixed points of $f(R)=\chi R^{n}$ and their  associated solutions. Here $a_0=H_1+H_2+H_3$. } \label{TavolaRn}
\begin{tabular}{llllll} \hline\hline
Point & Coordinates $\{\mathbb{R},\mathbb{K},\Omega\}$  & Scale Factor \\ \hline\\
$\mathcal{A}$ & $\left\{ 0, -1 , 0\right\}$  & \eqref{SolABC} \\ \\
$\mathcal{B}$ & $\left\{ 0, -1 ,-1- 3w\right\}$  & \eqref{SolABC}  \\ \\
$\mathcal{C}$ & $\left\{ n(1-n) , 2 (n-1) n-1, 0\right\}$  & \eqref{SolABC}  \\ \\
$\mathcal{D}$ & $\left\{ 0, 0 ,2- 3w\right\}$  & \eqref{SolDE}  \\ \\
$\mathcal{E}$ & $\left\{ 0,0, 0\right\}$  & \eqref{SolDE}  \\ \\
$\mathcal{F}$ & $\left\{ \frac{(5-4 n) n}{4 n^2-6 n+2}, 0, 0\right\}$&$a=a_0 (t-t_0)^{\frac{(1-2 n) (1-n)}{n-2}}$  \\ \\
$\mathcal{G}$ &
$\left\{ \frac{3-4 n+3 w}{4 n},0,\frac{6 n^2 w+8 n^2-9 n w-13 n+3 w+3}{2 n^2}\right\}$ &  $a=a_0 (t-t_0)^{\frac{2 n}{3 (w+1)}} $ \\ \\
\hline\hline\\
 \end{tabular}
   \end{center}
\end{table}
 
\begin{figure}[htbp]
\begin{center}
\includegraphics[scale=0.50]{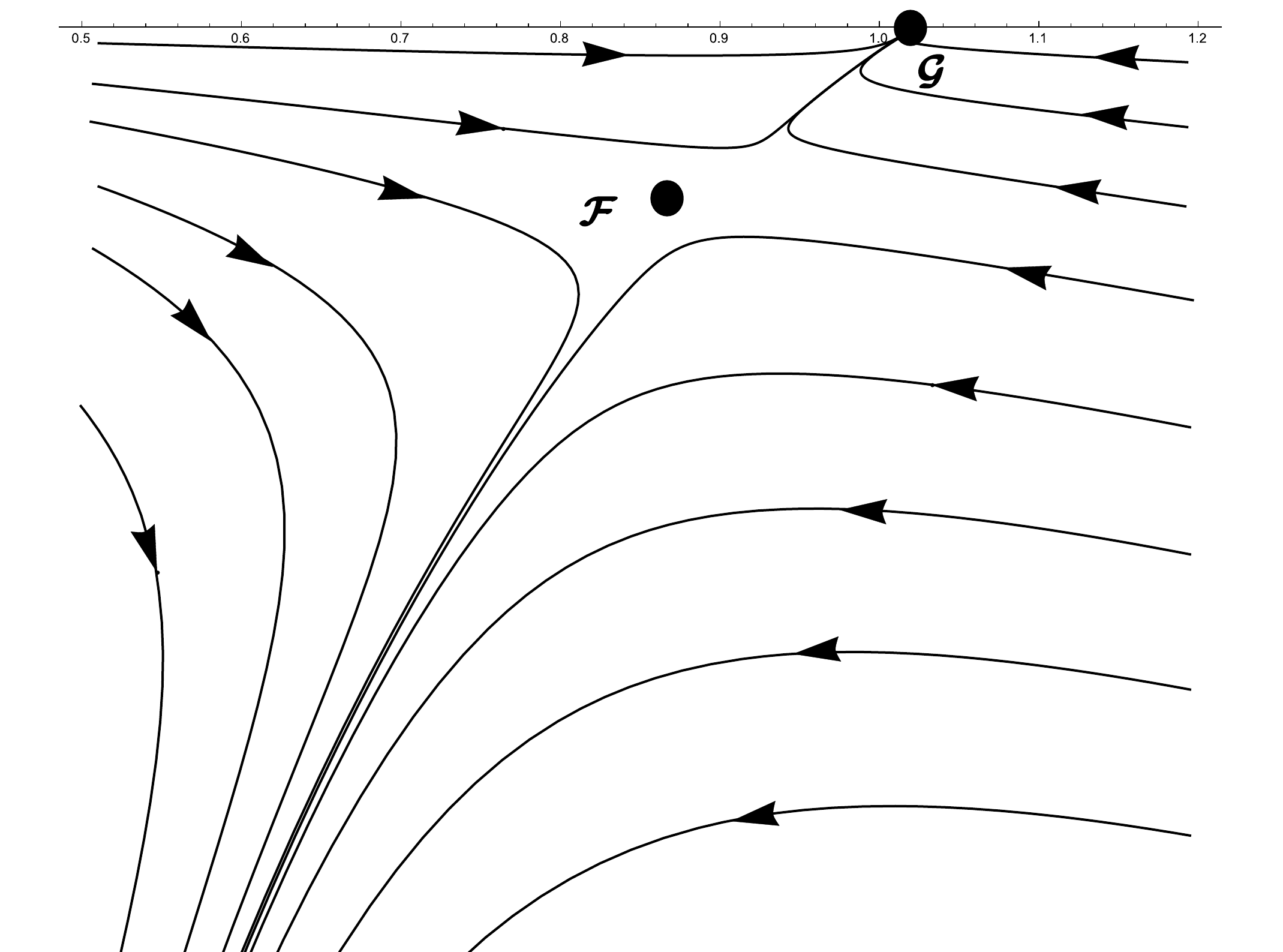}
\caption{Plot of a section of the invariant submanifold $K=0$ for the model $f(R)=\chi R^{n}$. Here the abscissa represents the variable $\mathbb R$ and the ordinate the variable $\Omega$.}
\label{FigRn}
\end{center}
\end{figure}

 \begin{table}[h]
\begin{center}
\caption{Stability of the fixed points of $f(R)=\chi R^{n}$  in the case $w=0$. Here A stays for attractor, R for repeller, S for saddle. } \label{TavolaRnStab}
 \begin{tabular}{lcccccccccccccc} \hline\hline
Point & $n<\frac{1}{2}\left(1-\sqrt{3}\right)$& $\frac{1}{2}\left(1-\sqrt{3}\right)<n<0$& $0<n<1/2$& $1/2<n<1$ 
\\ \hline\\
$\mathcal{A}$&  S & S & S& S \\ \\
$\mathcal{B}$ & S & S &S & S    \\ \\
$\mathcal{C}$ & S & A  & S &S \\ \\
$\mathcal{D}$&  R & R & R & R \\ \\
$\mathcal{E}$ & S & S &S & S    \\ \\
$\mathcal{F}$ & A & S  & S &A \\ \\
$\mathcal{G}$&  S & S & S& S \\ \\
\hline\hline\\
 \end{tabular}
  \begin{tabular}{lcccccccccccccc} \hline\hline
Point  & $1<n<5/4$& $5/4<n<4/3$& $4/3<n<\frac{1}{2}\left(1+\sqrt{3}\right)$& $n>\frac{1}{2}\left(1+\sqrt{3}\right)$
\\ \hline\\
$\mathcal{A}$& S& S& S& S   \\ \\
$\mathcal{B}$ &  S &S & S & S    \\ \\
$\mathcal{C}$ & S &S&A&S\\ \\
$\mathcal{D}$&  S & R & R& R \\ \\
$\mathcal{E}$ & S & S &S & S    \\ \\
$\mathcal{F}$ & R & S  & S &A \\ \\
$\mathcal{G}$&  S & S & S& S \\ \\

\hline\hline\\
 \end{tabular}
  \end{center}
\end{table}
It is instructive to compare the results above with the ones of \cite{Carloni:2004kp}. It is evident that the fixed points we have obtained, correspond to the ones  found in \cite{Carloni:2004kp}. In particular  the solutions associated to the fixed points are coincident when $H_2$ and $H_3$ are set to zero. This result shows how the original DSA would give  at best an incomplete result. For example in \eqref{SolABC}, since the $a^{-1}$ term is only relevant at small $a$, if $\mathcal A$, $\mathcal B$, $\mathcal C$ are attractors the cosmology will tend to become static after a phase of accelerated expansion. In terms of the stability, instead, the two methods  show a complete consistency: the nature of the fixed points is  the same as the one in \cite{Carloni:2004kp}. Specifically, comparing the ranges of stability the possibility of a transition between almost Friedmann (point $\mathcal{F}$)  and power law inflation/dark energy era (point $\mathcal G$)  is present also with the new method. In fact, since both the fixed points are on the $K=0$ invariant submanifold, we can check this result explicitly plotting this part of the phase space (see Figure \ref{FigRn}). It is evident that there is, in complete agreement with \cite{Carloni:2004kp}, a set of initial condition in which the transition appears.  The stability analysis results in Table \ref{TavolaRnStab} guarantees that this is the case also for orbits in the full phase space.

\subsection{The case $f(R)=R +\alpha R^{n}$}
Let us now consider another important model for fourth order gravity: the one in which a generic power of the Ricci scalar is added to the Hilbert-Einstein term. Historically this is the most studied form of fourth order gravity because of its appearance in several quantum gravity calculations \cite{BOS}. In cosmology the model with $n=2$ has gained particular attention as a model of geometric inflation \cite{Starobinsky:1980te}. 

The Lagrangian for this class of theories  can be written in the form showed in Section \ref{NewAction} as
\begin{equation}\label{fRRn}
f(R_0R,\alpha)=R_0 R+\alpha R_0^{n}R^{n}.
\end{equation}
Note that for $n<0$ this theory is not defined in $R=0$  and in $R_0=0$, but in fact we will see that, due to the presence of derivatives of $f$ with respect to $R$ in the cosmological equations (and therefore in the dynamical system) we will have divergences also in other intervals of $n$.

Substituting the $f$ above in the expression for ${\bf X}$, $\bf Y$ and $\bf Z$ we obtain
\begin{align}
\begin{split}
&{\bf X}=\frac{\mathbb{R}}{n}+\frac{ (n-1)\A \mathbb{R}^2}{n(\A \mathbb{R}
   +\alpha  n 6^{n-1}\A^{n} \mathbb{R}^{n})},\\
&{\bf Y}=\frac{4
   (n-1)}{\mathbb{R}}\left[1-\frac{\A \mathbb{R} }{\A \mathbb{R}
   +\alpha  n 6^{n-1}\A^{n} \mathbb{R}^{n}}\right],\\
&{\bf Z}=\frac{8
   (n-2) (n-1)}{3 \mathbb{R}^2}\left[1-\frac{\A \mathbb{R}}{\A \mathbb{R}
   +\alpha  n 6^{n-1}\A^{n} \mathbb{R}^{n}}\right].
\end{split}
\end{align}
Substituting in the general system \eqref{DynSysRed} we obtain
\begin{align}\label{DynSysRR^n}
\begin{split}
 &\DerN{\mathbb{R}}=\frac{\mathbb{R}
   \left[n\left(2 n-3 \right)\K-\left(2 n^2 +3 n+1\right)\mathbb{R} +n \Omega+4 n^2 -5 n\right]}{n(n-1) }\\
   &~~~~~~~~-\frac{(K-\Omega +1)}{6^{n-1} \alpha  (n-1) n \A^{n-1}  \mathbb{R}^{n-2}},\\
 &\DerN{\Omega}=\Omega\left[3 \K+\left(\frac{1}{n}-3\right)\mathbb{R}-\Omega +2-3 w-\frac{(n-1) \A \mathbb{R}^2}{ n(\A \mathbb{R}
   +\alpha  n 6^{n-1}\A^{n} \mathbb{R}^{n})}\right],\\
 &\DerN{\K}=2 \K(\K- \mathbb{R}+1),\\ 
   &\DerN{\mathbb{A} }=-2\mathbb{A} (2+\K-\mathbb{R}).
\end{split}
\end{align}
This system admits four invariant submainfolds ($\mathbb{K}=0, \Omega=0, \A=0, \mathbb{R}=0$). The last two invariant submanifolds can be singular, but  they can still contain some fixed points. As said, the presence of fixed points is based on the requirement of convergence of the cosmological equations and the eigenvalues of the fixed points. This implies that some fixed points will only exist for certain values of $n$.  

The list of the fixed points, their associated solutions and the interval of existence of the fixed points  are given in Table \ref{TavolaR+Rn}. Note that the points $\mathcal{A}-\mathcal{G}$ are the same of the ones found in the case of $R^n$-gravity. The presence of this type of fixed points can be understood thinking that nearby $\A=0$ the function $f$ can be approximated with its limit for small $R_0$. For the values of $n$ for which these fixed points exist the approximate function is $R_0^nR^n$ and therefore the fixed points of the case $f\propto R_0^nR^n$ appear also in the phase space of this theory.

On top of the points $\mathcal{A}-\mathcal{G}$ of the previous case we have some additional ones, which we will name $\mathcal{H}_i$. These points will exist  if $\alpha(n-2)>0$ (remember that the variable $\mathbb{A}$ is defined to be non negative) and their number depends on the value of $n$.
\begin{table}[h]
\begin{center}
\caption{Fixed points of $f(R)=R+\alpha R^{n}$ with their interval of existence and their associated solutions. } \label{TavolaR+Rn}
\begin{tabular}{llllll} \hline\hline
Point & Coordinates $\{\mathbb{R},\mathbb{K},\Omega, \A\}$  & Scale Factor & Existence
\\ \hline\\
$\mathcal{A}$ & $\left\{ 0, -1 , 0,0\right\}$  & \eqref{SolABC} & $n<1/2$ \\ \\
$\mathcal{B}$ & $\left\{ 0, -1 ,-1- 3w,0\right\}$  & \eqref{SolABC}& $n<1/2$   \\ \\
$\mathcal{C}$ & $\left\{ n(1-n) , 2 (n-1) n-1, 0,0\right\}$  & \eqref{SolABC} & $n<1$  \\ \\
$\mathcal{D}$ & $\left\{ 0, 0 ,2- 3w,0\right\}$  & \eqref{SolDE}& $n<1/2$   \\ \\
$\mathcal{E}$ & $\left\{ 0,0, 0,0\right\}$  & \eqref{SolDE}&  $n<1/2$   \\ \\
$\mathcal{F}$ & $\left\{ \frac{(5-4 n) n}{4 n^2-6 n+2}, 0, 0,0\right\}$&$a=a_0 (t-t_0)^{\frac{(1-2 n) (1-n)}{n-2}}$&  $n<1$ \\ \\
 \multirow{2}{*}{ $\mathcal{G}$}&$\left\{ \frac{3-4 n+3 w}{4 n},0\right.$ &   \multirow{2}{*}{ $a=a_0 (t-t_0)^{\frac{2 n}{3 (w+1)}} $}&   \multirow{2}{*}{ $n<1$}\\ 
 &$\left.,\frac{6 n^2 w+8 n^2-9 n w-13 n+3 w+3}{2 n^2},0\right\}$&&\\ \\
$\mathcal{H}_{i}$ &$\left\{2, 0, 0, 12 \sqrt[1-n]{\alpha(n-2)}\right\}$& \eqref{SolH} &$\alpha(n-2)>0$  \\  \\
\hline\hline\\
 \end{tabular}
   \end{center}
\end{table}

In  the $\mathcal{H}_i$ the scale factor is described by the equation
\begin{equation}\label{SolH}
\frac{\dot{a}}{a}=H_0+3 H_1\log a +9 H_2\log^2 a,
\end{equation}
which admits the exact solution
\begin{equation}\label{solH}
  a(t)=a_0 \exp \left\{\frac{\sqrt{4 H_2 H_0-H_1^2}}{2 H_2} \tan \left[\frac{1}{2} (t-t_0) \sqrt{4 H_2
   H_0-H_1^2}\right]-\frac{H_1}{2 H_2}\right\}.
\end{equation}
For $4 H_0 H_2-H_1^2>0$ this solution is monotonically growing with two inflection points at 
\begin{equation}
t=t^{*}_{1,2}=t_0\pm\frac{\arcsin\left(-\frac{\sqrt{4 H_0 H_2-H_1^2}}{3 H_2}\right)}{\sqrt{4 H_0
   H_2-H_1^2}}+2k\pi, \qquad k \in {\mathbb N}.
\end{equation}
and presents a discontinuity in
\begin{equation}
t=\bar{t}=t_0-\frac{\pi }{\sqrt{4 H_0 H_2-H_1^2}}+k\pi, \qquad k\in\mathbb{N}.
\end{equation}
For $t\rightarrow\bar{t}^+$ this solution approach to zero, whereas when $t\rightarrow\bar{t}^-$ the solution presents a vertical asymptote.

\begin{figure}%
\centering 
\subfigure[Plot of a period of \eqref{SolH} in the case $4 H_0 H_2-H_1^2>0$. Here $H_1=1$,  $H_2=3$, $H_3=2$  and $a_0=1$.]{\includegraphics[scale =0.4] {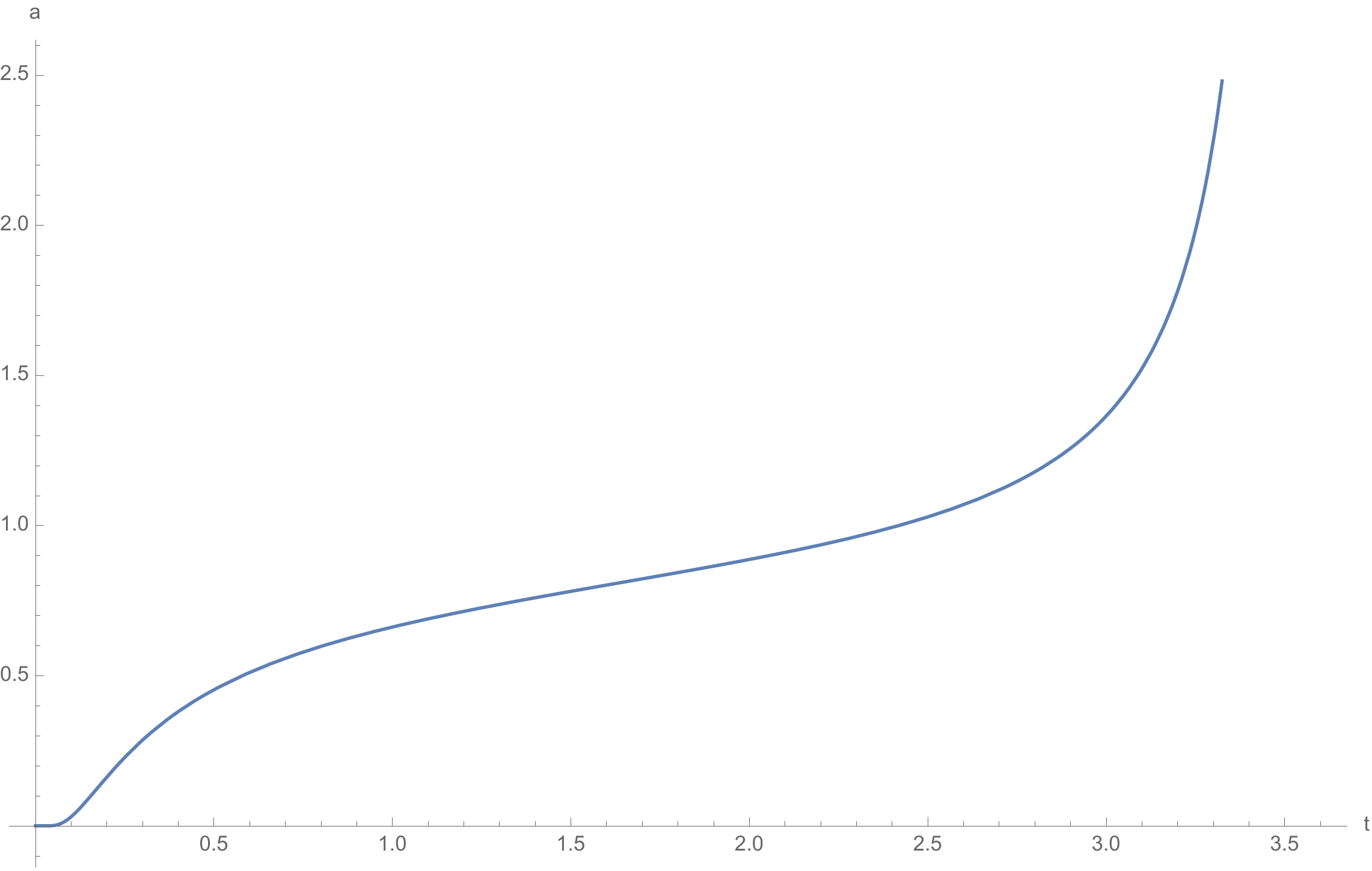}}\\
\subfigure[Plot of   \eqref{SolH} in the case $4 H_0 H_2-H_1^2<0$. The constants $H_i$ have all been chosen so that $P>0$ and $a_0=1$]{\includegraphics[scale =0.3] {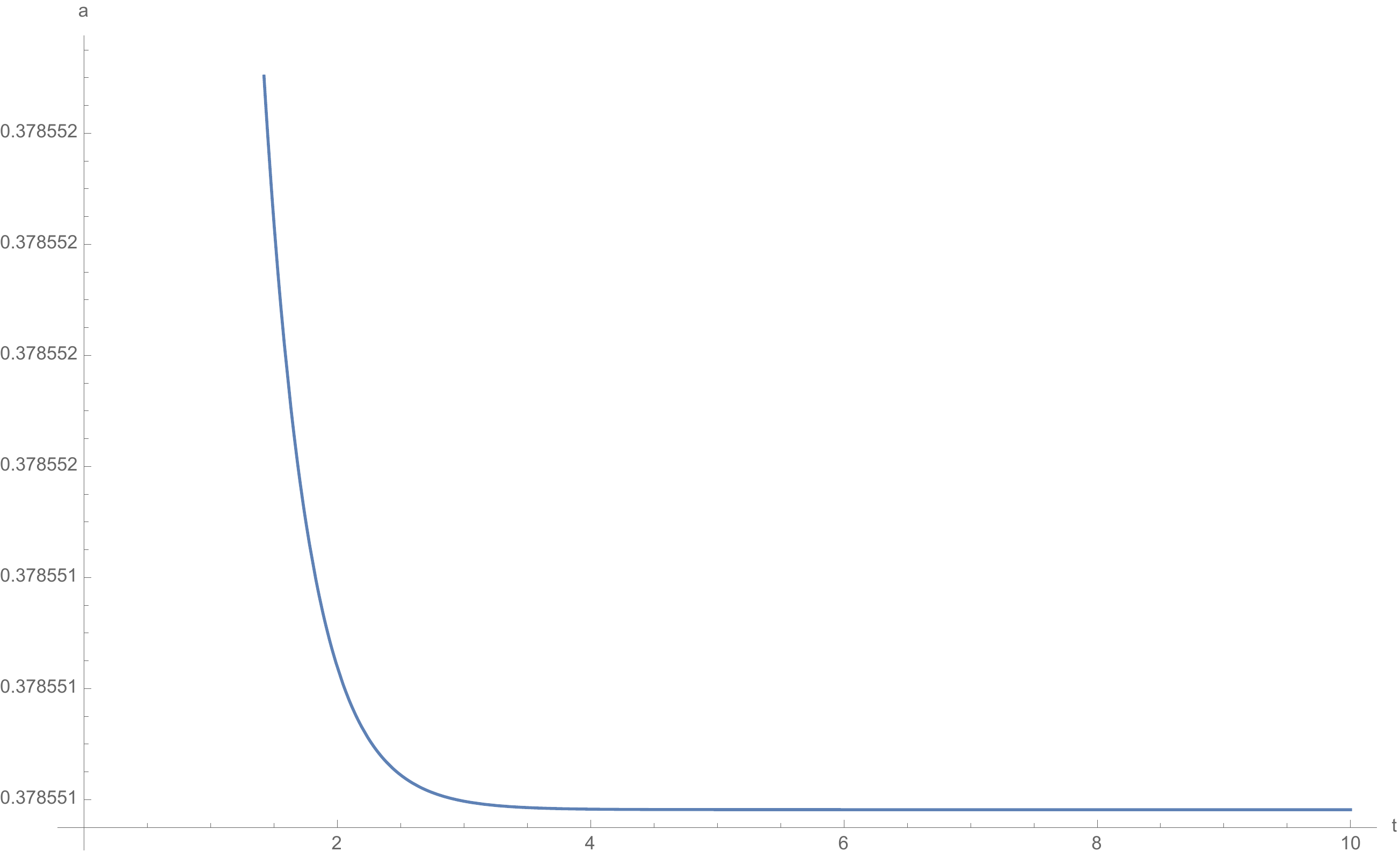}}\\ 
\subfigure[Plot of  \eqref{SolH} in the case $4 H_0 H_2-H_1^2<0$. The constants $H_i$ have all been chosen so that $P<0$ and $a_0=1$]{\includegraphics[scale =0.3] {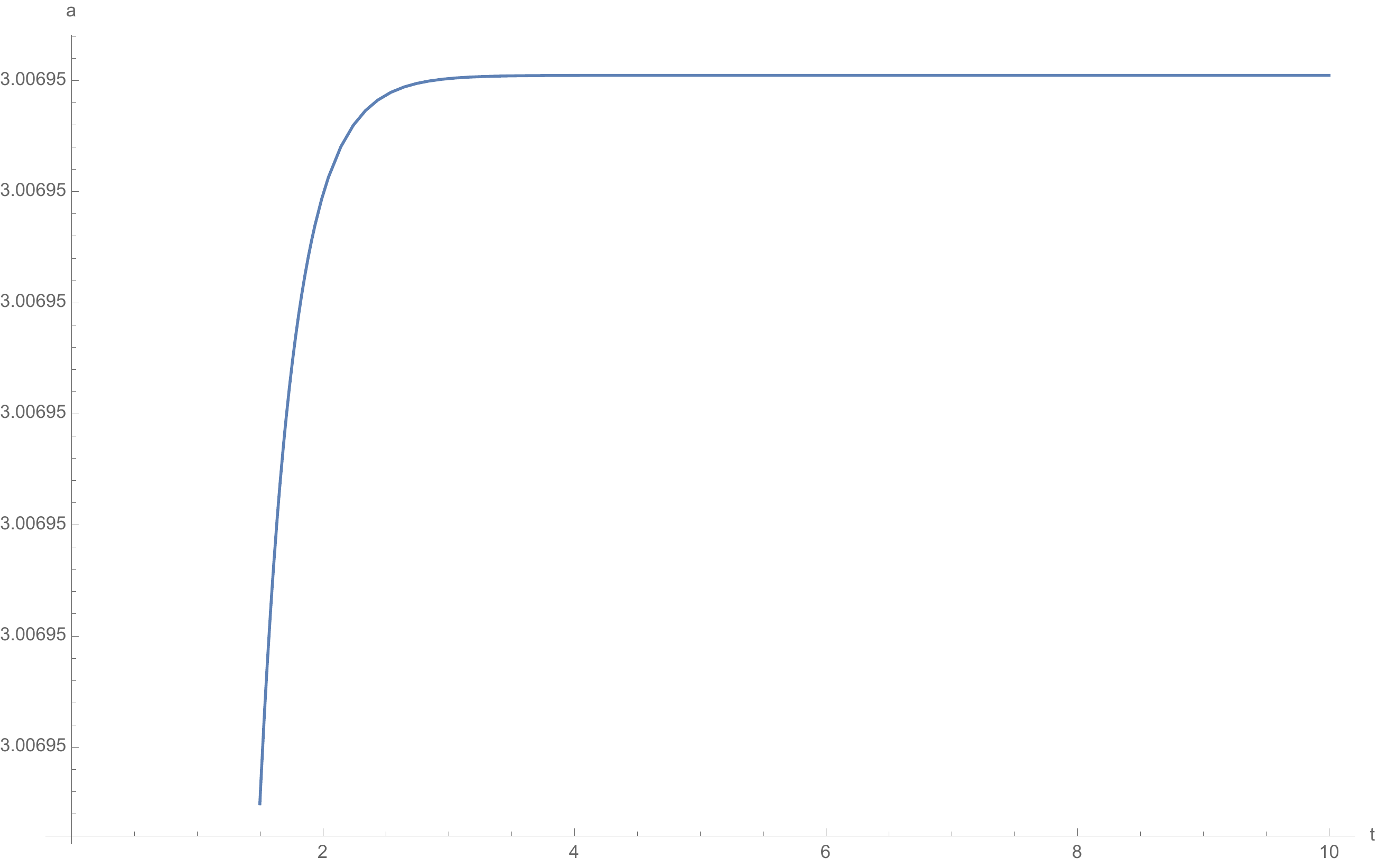}}\\ %
\caption{Plots of  \eqref{SolH} illustrating the different behaviour that this solution can represent.} \label{PlotSolH}
\end{figure}

In the case $4 H_0 H_2-H_1^2<0$ the \eqref{SolH} is instead not periodic and approaches a constant. The expanding or contracting character  depends on the sign of the quantity
\begin{equation}
P=-\frac{H_1+\sqrt{H_1^2-4 H_0 H_2}}{2 H_2},
\end{equation}
$P>0$ implies a growing scale factor and $P<0$ a decaying one. Plots of the solution \eqref{SolH} can be found in Figure \ref{PlotSolH}.

In the case $4 H_0 H_2-H_1^2>0$ the solution \eqref{solH} presents features which are physically very interesting. For times close to $\bar{t}^+$ the solution grows  exponentially, then around $t^{*}_1$ it changes into a decelerated expansion with non-constant deceleration factor. After a coasting phase around $t^{*}_2$, the decelerated expansion is followed by a new accelerated expansion phase. In a finite time (at $t=\bar{t}^-$), however, the solution becomes singular in the sense that $a$ and its derivatives as well as the Ricci scalar diverges at this specific time. This is a well known property of $f(R)$-gravity\footnote{These models in fact present also other types of singularities, like the ``weak singularities'' in \cite{Appleby:2009uf} or the ones found in \cite{Dolgov:2003px}.} \cite{Capozziello:2009hc,OdintsovSingularity}, but in this context one  is able to appreciate both this drawback of the theory and its potential as a model that unify inflation and dark energy.

Note that for $H_{1,2}=0$ the solution \eqref{solH} reduce to the standard de Sitter solution. This fact on one hand connects the points $\mathcal{H}_i$ to the de Sitter fixed point of the original DSA. On the other hand gives a hint of the true meaning of de Sitter solutions in the framework of $f(R)$-gravity. 

\begin{figure}[htbp]
\begin{center}
\includegraphics[scale=0.8]{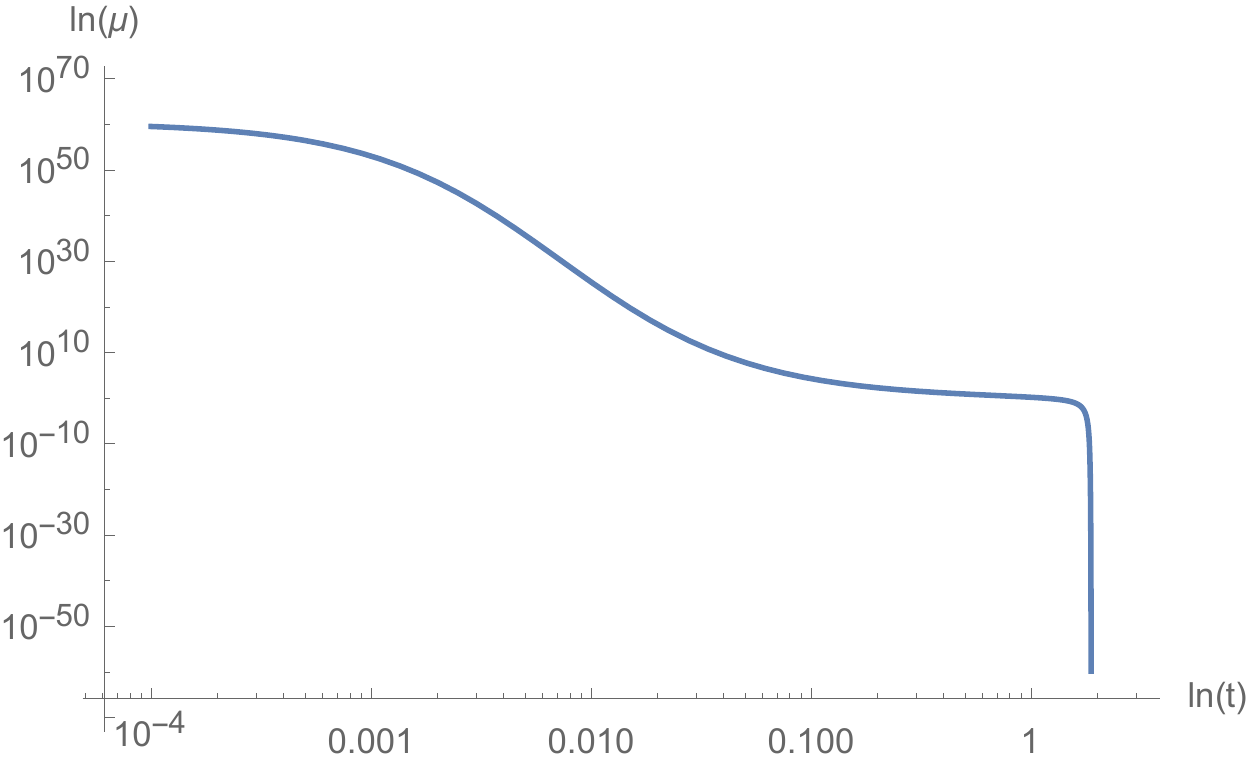}
\caption{Numerical solution for the energy density of \eqref{SolH} in the case $4 H_0 H_2-H_1^2>0$. Here $H_1=1$,  $H_2=3$, $H_3=2$  and $a_0=1$.}
\label{PlotSolH1_mu}
\end{center}
\end{figure}
\begin{figure}%
\centering 
\subfigure[Plot of the $\mathbb{R}>0$ section of the phase space in the case $n=1/4$ $\alpha=-10$.]{\includegraphics[scale =0.8] {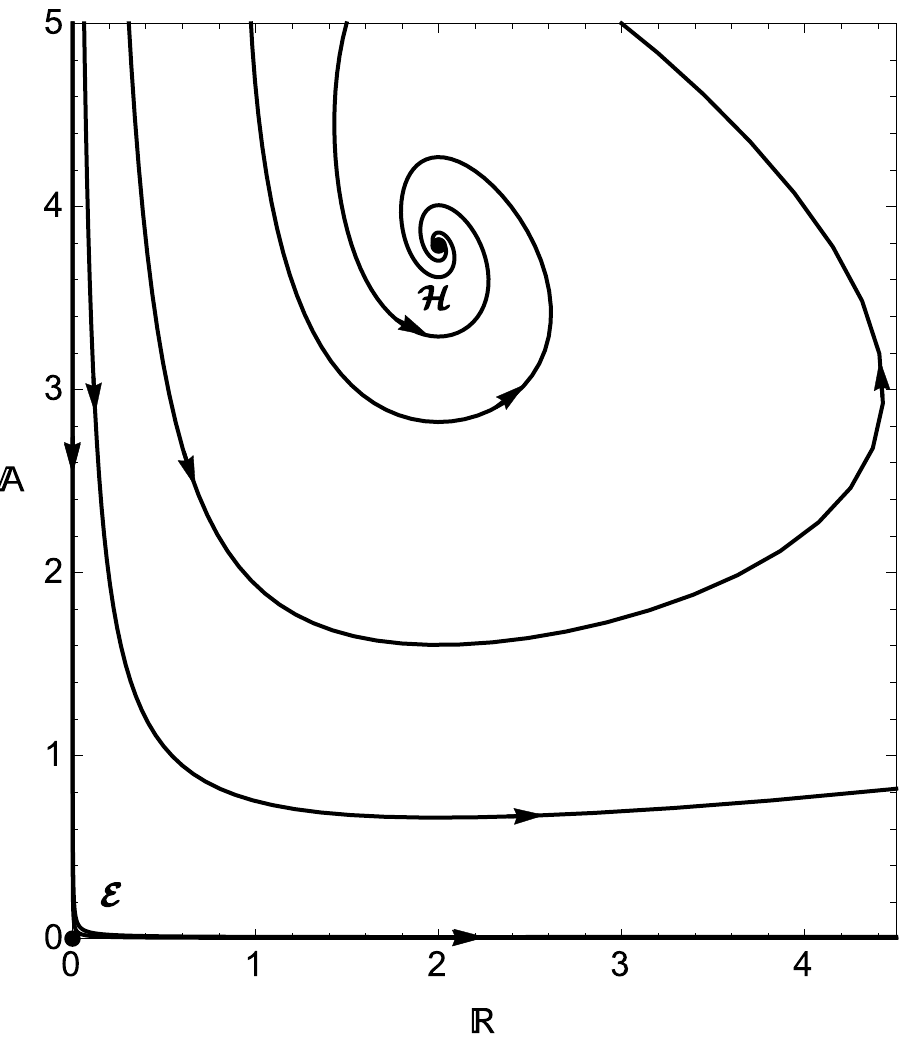}}\\
\subfigure[The case $n=3$ $\alpha=1/16$.]{\includegraphics[scale =0.8] {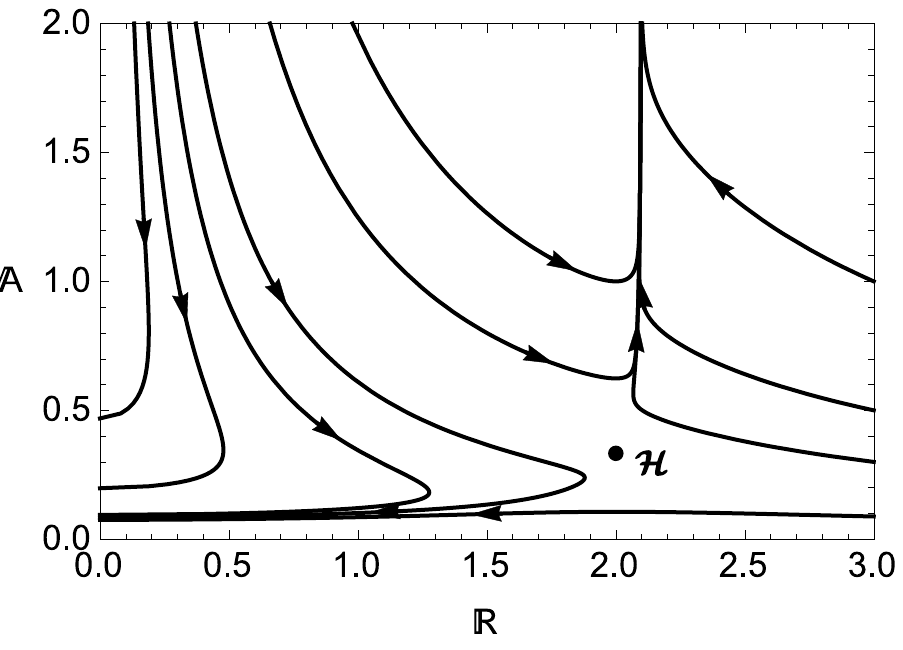}}\label{figRRn2}\\ 
\caption{Samples of the invariant submanifold $\Omega=0$, $\mathbb{K}=0$ for the theory $f(R)=R +\alpha R^{n}$. The values of the parameters have been chosen to give the best graphical representation of the phase space.} \label{PSRRn}
\end{figure}

The stability of the fixed points can be calculated as in the previous case using the Hartman Grobman theorem, and it is illustrated in Table \ref{TavolaRRnStab}. It is tempting to attempt a general derivation for the stability of points $\mathcal{H}_i$, however numerical inspection shows that such analysis might be unreliable. The calculations  show that  when points $\mathcal{H}_i$ exist there are only two options for their stability: they can be either a saddle or an attractor (or their focus counterpart). In Table \ref{TavolaRRnStab} we report some samples of the stability of points $\mathcal{H}_i$ for different values of the parameters $\alpha$ and $n$.  Examples of the phase space for these theories in the form of the invariant submanifold $\mathds{R},\A$ is given in Figures \ref{PSRRn}.
 \begin{table}[h]
\begin{center}
\caption{Stability of the fixed points of $f(R)=R+ \alpha R^{n}$  in the case $w=0$. Here  $n_1$ is the smallest real solution of the equation $256 n^3-608 n^2+417 n-81=0$ , A stays for attractor, R for repeller, S for saddle, F$_S$ for saddle focus.  The fixed points appear in the table only if they exist in at least one of the intervals of the parameter $n$ indicated.} \label{TavolaRRnStab}
 \begin{tabular}{lcccccccccccccc} \hline\hline
Point & $n<\frac{1}{2}\left(1-\sqrt{3}\right)$& $\frac{1}{2}\left(1-\sqrt{3}\right)<n<0$&  $0<n<n_1$&$n_1<n<1/2$ 
\\ \hline\\
$\mathcal{A}$&  S & S & S& S \\ \\
$\mathcal{B}$ & F$_S$ & F$_S$ &F$_S$ & F$_S$    \\ \\
$\mathcal{C}$ & S & S  & S &S \\ \\
$\mathcal{D}$&  S & S & S & S \\ \\
$\mathcal{E}$ & S & S &S & S    \\ \\
$\mathcal{F}$ & A & S  & S &S \\ \\
$\mathcal{G}$&  F$_S$ & F$_S$ & F$_S$& S \\ \\
\hline\hline\\
 \end{tabular}
  \begin{tabular}{lcccccccccccccc} \hline\hline
Point  &$1/2<n<1$ &$n>1$
\\ \hline\\
$\mathcal{C}$ & S & NA 
\\ \\
$\mathcal{F}$ & S &  NA 
\\ \\
$\mathcal{G}$&  F$_S$ &  NA  
\\ \\
\hline\hline
\end{tabular}
\end{center}
\end{table}
 \begin{table}[h]
\begin{center}
\caption{Stability of the fixed points $\mathcal{H}_i$ of $f(R)=R+ \alpha R^{n}$  in the case $w=0$. Here A stays for attractor, S for saddle, F$_A$ for saddle focus and NA represents the absence of fixed points. In the cases considered there is only one fixed point $\mathcal H$. } \label{TavHRRn}
 \begin{tabular}{l|cccccccccccccc} \hline\hline
&$n=-2$ &$n=-1$ & $n=3/2$  & $n=5/2$
\\ \hline\\
$\alpha=-2$& S & S & NA& NA \\ \\
$\alpha=-1$ & S & S & NA & NA  \\ \\
$\alpha=1$ & A & A  & NA &NA \\ \\
$\alpha=2$&  NA & NA & S & S \\
\hline\hline
\end{tabular}
  \end{center}
\end{table}
Since this model can be treated also with the original DSA it is useful to make a comparison between the results we obtained above and the ones in \cite{Carloni:2007br}. Differently from the case of $R^n$ gravity the phase space obtained by the two methods is not the same. In particular, the original DSA returns a phase space with many more fixed points. The origin of this difference is probably to be attributed to the choice of variables of the original DSA. For the common points one can compare the results on the stability and it is easy to verify complete consistency. For example, the de Sitter solution that in \cite{Carloni:2007br} is associated with $\mathcal{E}^*$ has a stability that coincide with the point $\mathcal{H}$ of the present analysis. 

In the case $n=2$ the new DSA returns a phase space with no finite fixed points. It is known that the case $n=2$ in the theory \eqref{fRRn} presents significant physical differences with respect to the other model of this class \cite{BOS}, and it is to be expected that the phase space will reflect these differences. The fact that the phase space does not present a point of type  $\mathcal{H}$ does not necessarily imply that the model does not have de Sitter solutions (we know in fact that  they are present \cite{Barrow:1983rx}). A careful analysis of the  equations shows that the fixed point in this case is asymptotic  ($\mathbb{A}\rightarrow\infty$) and it is therefore excluded by the present analysis. 

It is interesting to note that one of the conclusions in \cite{Capozziello:2009hc} is that to avoid the singularity one either has to recur to special initial conditions or to add additional curvature invariants. This result seems consistent with our findings. In order to avoid the singularity one either has to control the initial conditions (by setting to zero some of the constant $H_i$), the values of the parameters like in Figure \ref{figRRn2}, or hope that adding additional curvature invariants the fixed points $\mathcal{H}$ will become irrelevant (in the same way of what happens with the case $n=2$ above). The issue is that the feature of the inflation dark energy connection seems to be inextricably tied to the approach to the singularity. Thus, in order to generate cosmic acceleration without incurring in the singularity requires that other dynamical mechanisms/forms of the function $f$ have to be found.

\subsection{The Starobinsky model.}
The Starobinsky model is one of the most important class of models of Dark Energy based on fourth order gravitation \cite{Starobinsky:2007hu}. The basic idea is to construct, via a function of the Ricci scalar, an effective cosmological constant which is relevant in curved spacetimes, but approaches zero in the case of flat spacetime.  The function $f$ of this model is given by
\begin{equation}\label{StarL}
f(R_0,R)=R+\lambda \bar{R}_0\left[\left(1+\frac{R^{2}}{\bar{R}_0^{2}}\right)^{-n}-1\right],
\end{equation}
with $n$ and $\lambda$ positive and $\bar{R}_0$ of the order of the inverse of the present value of the cosmological constant. The parameter $\lambda$ and the value of $n$ are related by an algebraic equation. In \cite{Starobinsky:2007hu} considerations on the Solar System constraints and cosmological linear perturbation theory were used to find that one should have $n\geq2$ and $\lambda\gtrsim 0.94$.

Following Section \ref{NewAction} we can write the Lagrangian \eqref{StarL} for this class of models as
\begin{equation}\label{Starf}
f(R)=R_0 R+\alpha \left[\left(1+\beta^2 R_0^{2}R^{2}\right)^{-n}-1\right],
\end{equation}
where the parameters $\alpha,\beta$ are related to the ones in \eqref{StarL} by the relations $\lambda=\alpha \left|\beta \right|R_0$ and $\bar{R}_0=\left(\left|\beta\right| R_0\right)^{-1}$.

The functions $\mathbf X$, $\mathbf Y$, $\mathbf Z$ are
\begin{align}\label{XYZ_Star}
\begin{split}
&{\mathbf X}=\frac{\alpha +12 \alpha \beta  \mathbb{A}  \mathbb{R} [3 \mathbb{A}\mathbb{R} (1+2 n)-\alpha  n]}{6 \mathbb{A} \left[\left(1+ 6^2 \mathbb{A}^2 \beta  \mathbb{R}^2\right)^{n+1}-12 \alpha
    \mathbb{A} \beta  n \mathbb{R}\right]}+\frac{6 \mathbb{A} \mathbb{R}-\alpha }{6 \mathbb{A}},\\
   &{\mathbf Y}=\frac{3\, 2^4 \alpha \beta  n  \mathbb{A} \left(36 \mathbb{A}^2 \beta  (2 n+1) \mathbb{R}^2-1\right)}{\left(1+6^2\beta \mathbb{A}^2 \mathbb{R}^2\right) \left[\left(1+6^2 \mathbb{A}^2 \beta  \mathbb{R}^2\right)^{n+1}-12 \alpha
    \mathbb{A} \beta  n \mathbb{R}\right]},\\
   &{\mathbf Z}=-\frac{2^8 3^3\alpha  \beta ^2 n (n+1) \mathbb{A}^5\mathbb{R} \left(12 \mathbb{A}^2 \beta  (2
   n+1) \mathbb{R}^2-1\right)}{\left(1+6^2\beta \mathbb{A}^2 \mathbb{R}^2\right)^2 \left[\left(1+6^2 \mathbb{A}^2 \beta  \mathbb{R}^2\right)^{n+1}-12 \alpha
    \mathbb{A} \beta  n \mathbb{R}\right]},
\end{split}
\end{align}
and the dynamical system  \eqref{DynSysRed}  becomes
\begin{align}\label{DynSysStar}
  \begin{split}
 &\DerN{\mathbb{R}}=-\frac{1}{2 6^2 n \mathbb{A}^2\left(2 n 6^2\mathbb{A}^2 
   \mathbb{R}^2+6^2 \mathbb{A}^2 \mathbb{R}^2-1\right)}\left\{1+2\, 6^2  \beta \A^2 \mathbb{R} [n (\K-\mathbb{R}+\Omega +3)+\mathbb{R}]\right.\\
   &~~~~~~~~\left.6^4  \beta ^2 \mathbb{R}^3 \mathbb{A}^4 [2 n (3 \mathbb{R}+\Omega-\K (4 n+3)+4 n (\mathbb{R}-2)
   -5)+\mathbb{R}]\right\}\\
   &~~~~~~~~-\frac{[6 \A (\K-\Omega +1)-\alpha ] \left(6^2 \A^2 \beta  \mathbb{R}^2+1\right)^{n+2}}{2 6^2 \alpha   \beta  n \mathbb{A}^2\left[36 \A^2 \beta  (2 n+1) \mathbb{R}^2-1\right]},\\
 &\DerN{\Omega}=\frac{\alpha  \Omega  \{12  \beta \A  \mathbb{R} [3 \A \mathbb{R}(2 n +1)-\alpha  n]+1\} }{6 \mathbb{A} \left[\left(1+ 6^2 \mathbb{A}^2 \beta  \mathbb{R}^2\right)^{n+1}-12 \alpha
    \mathbb{A} \beta  n \mathbb{R}\right]}\\
   &~~~~~~~~-\frac{\Omega  \{\mathbb{A}[6(2+3w)-18  \K+12
   \mathbb{R}+6 \Omega]+\alpha \}}{6 \mathbb{A}},\\
 &\DerN{\K}=2 \K(\K- \mathbb{R}+1),\\
   &\DerN{\mathbb{A} }=-2\mathbb{A} (2+\K-\mathbb{R}).
\end{split}
\end{align}
The system presents three invariant submanifolds: $K=0$, $\Omega=0$, $\mathbb{A}=0$, although the last submanifold is singular. Note that, differently form the previous case, none of the fixed points $\mathcal{A}-\mathcal{G}$ is present here. This can be explained looking at the limit of the action for $R_0\rightarrow 0$. In this case in fact the action reduces to $R_0 R$ rather than $R_0^n R^n$ and therefore the fixed points of $R^n$-gravity are not present.

Even if this system is valid for any value of the parameters, it is not possible to find its fixed points analytically\footnote{Of course the calculation could be done numerically, but we do not perform such task here.}: this task would entail the resolution of an algebraic equation of order $n$ for which no general analytical solution is known. We therefore refer to the bounds in \cite{Starobinsky:2007hu} and we set from now on $n=2$ and $\beta,\alpha=1$. In this case the system admits three  fixed points with real coordinates (see Table \ref{TavolaStar}). One of them is on the singular submanifold and presents singular eigenvalues so that it does not appear in Table \ref{TavolaStar}. The other two ($\mathcal{H}_1$ and $\mathcal{H}_2$) are both associated to a solution of the type \eqref{SolH}. 

The stability analysis reveals that the character of the points $\mathcal H_i$ is in general different. In our particular case, only one of these point is  an attractor whereas the other is unstable. Since the phase space contains invariant submanifolds, we can conclude that for these values of the parameters there is only a specific set of initial conditions which lead to the attractor $\mathcal{H}_2$. Therefore, also in the case of the Starobinsky model there is the possibility that the cosmology will evolve towards a singularity, but this occurrence depends strictly on the choice of the initial conditions. In orbits which do not approach $\mathcal H$, the fate of the cosmological models depends on the presence of asymptotic attractors.

\begin{table}[h]
\begin{center}
\caption{Fixed points for the Starobinsky model and their stability in the case $n=2$ and $\alpha=\beta=1$. Here $\A_{0}$ is the only positive real solution of the equation $(6 \A-1) \left(144 \A^2+1\right)^4+144 \A^2 \left(432 \A^2+4\right)+1=0$ different form $1/12$, S represents a saddle and F$_A$ is an attractive focus.} \label{TavolaStar}
\begin{tabular}{llllll} \hline\hline
Point & Coordinates $\{\mathbb{R},\mathbb{K},\Omega, \A\}$  & Scale Factor & Stability
\\ \hline\\
$\mathcal{H}_1$&$\left\{2, 0, 0, \frac{1}{12}\right\}$&\eqref{SolH}& S \\ \\
$\mathcal{H}_2$&$\left\{2, 0, 0, \A_{0}\right\}$&\eqref{SolH}& F$_A$  \\ \\
\hline\hline\\
 \end{tabular}
   \end{center}
\end{table}

\begin{figure}[htbp]
\begin{center}
\includegraphics[scale=0.8]{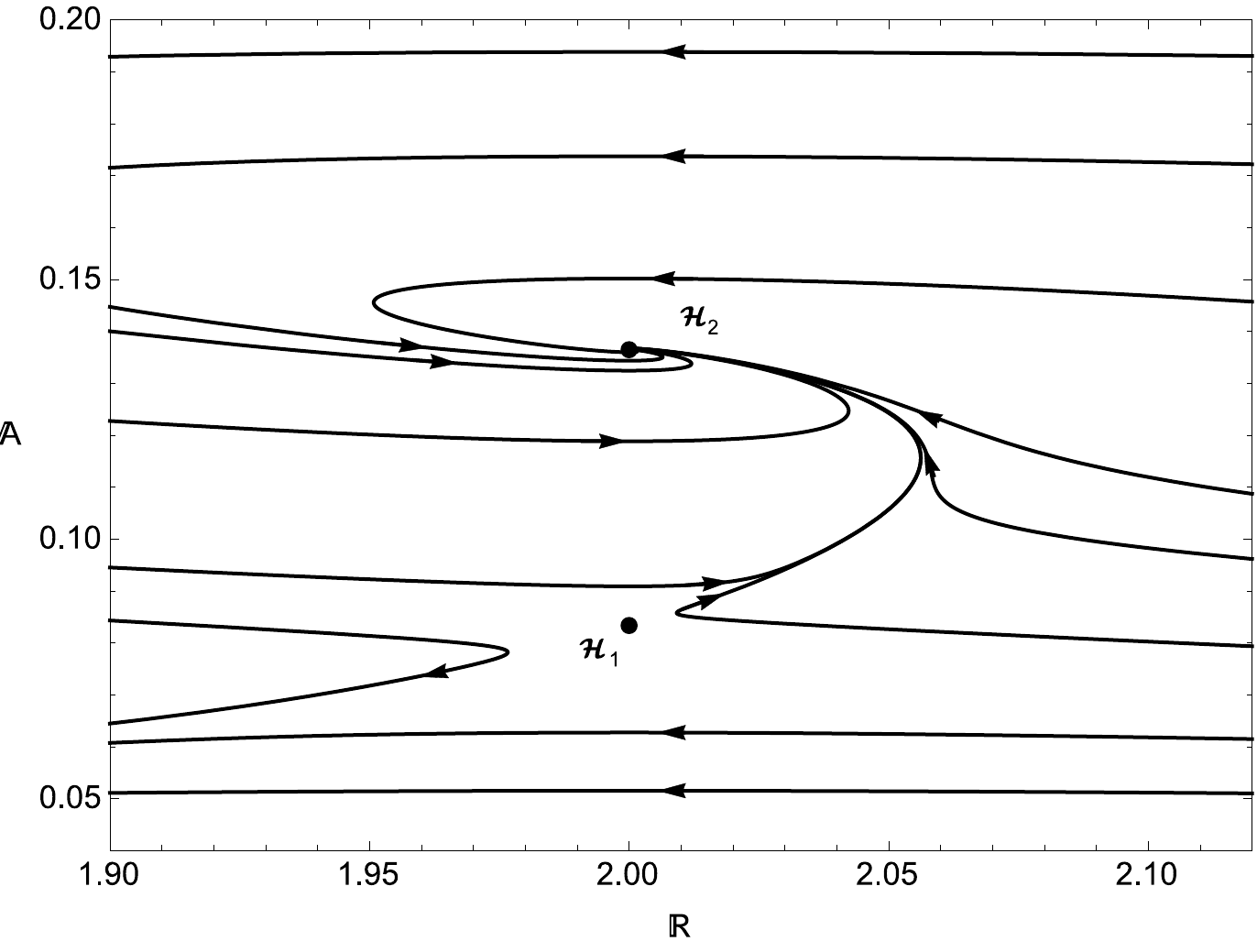}
\caption{The section of the invariant submanifold $\K=0,\Omega=0$ of the phase space of the Starobinsky model which contains $\mathcal{H}_1$ and $\mathcal{H}_2$ in the case $n=2$, $R_0>0$ and $\beta,\alpha=1$.}
\label{PSStar}
\end{center}
\end{figure}

A plot of the section of the invariant submanifold $\K=0,\Omega=0$ which contains $\mathcal{H}_1$ and $\mathcal{H}_2$ is given in Figure \ref{PSStar}.

\subsection{The Hu-Sawicki model}
As a last example we consider a model for geometric DE proposed by Hu and Sawicki \cite{Hu:2007nk}. The model was designed to be able to reproduce cosmic acceleration without the explicit introduction of a cosmological constant and, at the same time,  to be compatible with cosmological and Solar System tests.

The action for the Hu-Sawicki model can be written as \eqref{lagr f(R)R0} setting
\begin{equation}\label{HSf}
f(R_0,R)=R_0 R-\frac{\alpha R_0^n R^n}{1+\beta R_0^n R^n},
\end{equation}
where the parameter $n$ is chosen to be positive, $\alpha>0$ and $\beta>0$. With this choice of $f$ the functions $\mathbf X$, $\mathbf Y$, $\mathbf Z$ are given by
\begin{align}\label{XYZ_HS}
  \begin{split}
&{\mathbf X}=\frac{6 \beta \mathbb{A}   \mathbb{R}-\alpha}{6 \mathbb{A} \beta }+\frac{\alpha\, 6^2 \mathbb{A}^2 \mathbb{R}^2 \left[\beta  6^{n} \mathbb{A}^{n}
   \mathbb{R}^{n}(1+n) -\alpha n  6^{n-1} \mathbb{A}^{n-1} \mathbb{R}^{n-1}+1\right]}{6 \mathbb{A} \mathbb{R}+6^n
   \mathbb{A}^n \mathbb{R}^n (12 \mathbb{A} \beta  \mathbb{R}-\alpha  n)+\beta ^2 6^{2 n+1} \mathbb{A}^{2 n+1} \mathbb{R}^{2 n+1}},\\
   &{\mathbf Y}=-\frac{4 n \left[ 12 \mathbb{A} \mathbb{R}-6^n \mathbb{A}^n \mathbb{R}^n (\alpha -12 \mathbb{A} \beta 
   \mathbb{R}+\alpha  n)\right]}{6 \mathbb{A} \mathbb{R}^2 \left[6 \mathbb{A} \mathbb{R}+6^n
   \mathbb{A}^n \mathbb{R}^n (12 \mathbb{A} \beta  \mathbb{R}-\alpha  n)+\beta ^2 6^{2 n+1} \mathbb{A}^{2 n+1} \mathbb{R}^{2 n+1}\right]}\\
   &~~~~~~~-\frac{8 n}{\mathbb{R} \left(\beta  6^n \mathbb{A}^n \mathbb{R}^n+1\right)},\\
   &{\mathbf Z}=\frac{2^3 n 6^2 \mathbb{A} \mathbb{R} \left[\beta  6^{n} \mathbb{A}^{n} \mathbb{R}^{n}(n+1)-\alpha  6^{n-2}
    \mathbb{A}^{n-1} \mathbb{R}^{n-1}(n^2+2+3n)+1\right]}{3
   \mathbb{R}^2 \left[6 \mathbb{A} \mathbb{R}+6^n
   \mathbb{A}^n \mathbb{R}^n (12 \mathbb{A} \beta  \mathbb{R}-\alpha  n)+\beta ^2 6^{2 n+1} \mathbb{A}^{2 n+1} \mathbb{R}^{2 n+1}\right]}\\
   &~~~~~~~~+\frac{16n^2}{\mathbb{R}^2 \left(\beta  6^n \mathbb{A}^n \mathbb{R}^n+1\right)^2}-\frac{16 n \left(n+1\right)}{\mathbb{R}^2 \left(\beta  6^n \mathbb{A}^n
   \mathbb{R}^n+1\right)}.
\end{split}
\end{align}
Substituting in \eqref{DynSysRed}, we obtain the dynamical system  
\begin{align}\label{DynSysHS}
  \begin{split}
 &\DerN{\mathbb{R}}=\frac{\mathbb{R}}{\alpha  n (n+1)^2} \left\{\alpha n  (3 \K-\Omega
   +5)+\mathbb{R} [\alpha -12 \mathbb{A} \beta  (\K-\Omega +1)]+2 \alpha  n^3 (\K-\mathbb{R}+2)\right. \\
   &~~~~~~~~\left.+\alpha  n^2 (5 \K-5 \mathbb{R}-\Omega +9)-24 n \mathbb{A} \beta  \mathbb{R} (\K-\Omega +1)\right\}\\
   &~~~~~~~~+\frac{2 n \mathbb{R} \{[24 \mathbb{A} \beta  n \mathbb{R} -\alpha \K (n^2-1 )] (\K-\Omega+1)+\alpha (n-1)^2
   \mathbb{R}\}}{\alpha  \beta  6^n (n-1) (n+1)^3 \mathbb{A}^n
   \mathbb{R}^n-\alpha  \left(n^2-1\right)^2}\\
   &~~~~~~~~+\frac{\beta  6^n \mathbb{A}^n \mathbb{R}^{n+2} [\alpha-6 \mathbb{A} \beta( 1+ \Omega  + 
 \K)]}{\alpha  n (n+1)}+\frac{6^{1-n} \mathbb{A}^{1-n}  \mathbb{R}^{2-n}}{\alpha n
   (n-1)}(\K-\Omega +1),\\
 &\DerN{\Omega}=\frac{ \alpha \mathbb{R} \Omega  \left[(\beta (n+1) - n\alpha) 6^{n} n
   \mathbb{A}^{n} \mathbb{R}^{n}+1\right]}{6 \beta \mathbb{A}\mathbb{R}\left( \beta \, 6^{n} \mathbb{A}^{n} \mathbb{R}^{n}+1\right)^2-\alpha \beta  n  6^{n} \mathbb{A}^{n} \mathbb{R}^{n}}-\Omega\left(\Omega -3  \K+2  \mathbb{R}+  3w-2+\frac{\alpha }{6 \mathbb{A}\beta }\right),\\
 &\DerN{\K}=2 \K(\K- \mathbb{R}+1),\\
   &\DerN{\mathbb{A} }=-2\mathbb{A} (2+\K-\mathbb{R}),
\end{split}
\end{align}
This system can present four invariant submanifolds (${\mathbb K} =0$, $\Omega=0$, ${\mathbb A} =0$ and ${\mathbb R} =0$), but the last one exists only for $0<n<2$. The  ${\mathbb A} =0$ submanifold can be singular. This happens for $n>1$.   Differently from the Starobinsky model, the phase space in this case contains different fixed points depending on the value of the parameter $n$. In particular, the conditions $\beta>0$ and $\mathbb{A}>0$ limits strongly the number of fixed point belonging to the class $\mathcal{H}$. The list of fixed points for $n\neq1$  is given in Table \ref{TavolaHS}.   The case $n=1$ requires a special treatment and will be covered in a separate subsection. 

As usual, the conditions of existence for the fixed points in Table \ref{TavolaHS} have been determined asking that  the dynamical system should be at least of class $C(2)$ in the fixed point. In one case  (the point $\mathcal{C}$) there are values of $n$ for which the Jacobian is divergent, but the eigenvalues converge. We refer as interval of existence of the point $\mathcal{C}$ the one of convergence of the eigenvalues. Within the interval of existence of the fixed points the stability can be determined with the standard methods. The results are given in Table \ref{TavolaHSStab}. As in the case of $f(R)=R+\alpha R^n$ the stability of  $\mathcal{H}_i$ needs to be calculated case by case, but it can only be a saddle or an attractor of their focus counterparts. An example of the stability of points $H_i$ for different combinations of the values of $\alpha$, $\beta$ and $n$ in the case of dust ($w=0$) is given in Table \ref{TavolaHSStab}

In this case the points  $\mathcal C$  and $\mathcal H$ are the only possible finite attractors for the cosmology. The first point, however, only exists for $ 0<n<1$. Therefore for $n>1$, we are in the same situation of the Starobinsky model: only a careful choice of the initial conditions could avoid the singularity of solution \eqref{SolH}. 

\begin{table}[h]
\begin{center}
\caption{Fixed points of the Hu-Sawicki model for $n\neq1$ and their associated solutions. Here $\A_0$ represents the positive solutions of the equation given in the last line of the Table. } \label{TavolaHS}
\begin{tabular}{lllll} \hline\hline
Point & Coordinates $\{\mathbb{R},\mathbb{K},\Omega, \A\}$  & Scale Factor & Existence
\\ \hline\\
$\mathcal{A}$ & $\left\{ 0, -1 , 0,0\right\}$  & \eqref{SolABC} & $0<n<1$ \\ \\
$\mathcal{B}$ & $\left\{ 0, -1 ,-1- 3w,0\right\}$  & \eqref{SolABC}& $0<n<1$   \\ \\
$\mathcal{C}$ & $\left\{ n(1-n) , 2 (n-1) n-1, 0,0\right\}$  & \eqref{SolABC} & $0<n<1$  \\ \\
$\mathcal{D}$ & $\left\{ 0, 0 ,2- 3w,0\right\}$  & \eqref{SolDE}& $0<n<1$   \\ \\
$\mathcal{E}$ & $\left\{ 0,0, 0,0\right\}$  & \eqref{SolDE}&  $0<n<1/2$   \\ \\
$\mathcal{F}$ & $\left\{ \frac{(5-4 n) n}{4 n^2-6 n+2}, 0, 0,0\right\}$&$a=a_0 (t-t_0)^{\frac{(1-2 n) (1-n)}{n-2}}$&  $0<n<1$ \\ \\
$\mathcal{H}_i$ &$\left\{2, 0, 0, \mathbb{A}_0\right\}$& \eqref{SolH} & \\ \\
\hline\\ \\
\multicolumn{5}{c}{$\frac{12^{-n} \mathbb{A}^{-n} \left\{12 \mathbb{A}+2^{6 n+1} \beta ^2 27^n \mathbb{A}^{3 n} (6 \mathbb{A} \beta
   -\alpha )+\beta  144^n \mathbb{A}^{2 n} [36 \mathbb{A} \beta +\alpha 
   (n-4)]+12^n \mathbb{A}^n [36 \mathbb{A}\beta +\alpha  (n-2)]\right\}}{\alpha  n
   \left[\beta  12^n (n+1) \mathbb{A}^n-n+1\right] }=0$}\\ \\
   \hline\hline\\
 \end{tabular}
   \end{center}
\end{table}

 \begin{table}[h]
\begin{center}
\caption{Stability of the fixed points of the Hu-Sawicki model  in the case $n\neq1$ and $\alpha\neq 1$. Here A stays for attractor,  S for saddle and  F$_A$ for attractive focus. The value  of the $\A$ coordinate of $\mathcal{H}$ is approximated.} \label{TavolaHSStab}
 \begin{tabular}{lcccccccccccccc} \hline\hline
Point & $0<n<1/2$& $1/2<n<1$& $n>1$ 
\\ \hline\\
$\mathcal{A}$& S&S&NA \\ \\
$\mathcal{B}$ &S &S& NA \\ \\
$\mathcal{C}$ & F$_A$ &F$_A$&NA\\ \\
$\mathcal{D}$&  S&S&NA\\ \\
$\mathcal{E}$ &  S&NA&NA\\ \\
$\mathcal{F}$ & S&S&NA\\ \\
$\mathcal{G}$&  S &S& NA\\ \\
\hline\hline\\
 \end{tabular}
  \begin{tabular}{l|cccccccccccccc} \hline\hline
$(n,\alpha,\beta)$ & Coordinates of $\mathcal{H}_i$  & Stability
\\ \hline\\
\multirow{2}{*}{$(2,3, 1/2)$}&  $\left\{2, 0, 0, 0.07\right\}$ & S  \\
&  $\left\{2, 0, 0, 0.97\right\}$ & F$_A$  \\ \\
\multirow{2}{*}{$(3,4, 7/10)$}&  $\left\{2, 0, 0, 0.89\right\}$ & F$_A$  \\
&  $\left\{2, 0, 0, 0.95\right\}$ &  S \\ \\
\multirow{2}{*}{$(4,3, 3/5)$}&  $\left\{2, 0, 0, 0.10\right\}$ & S  \\
&  $\left\{2, 0, 0, 0.83\right\}$ & F$_A$  \\ \\
\hline\hline
\end{tabular}
  \end{center}
\end{table}

 \subsubsection{The case $n=1$.}
It is interesting to note that the dynamical system \eqref{DynSysHS} is not defined for $n=1$, which means that the dynamical system in this case needs to be obtained re-deriving  $\mathbf X$, $\mathbf Y$, $\mathbf Z$ and the equations \eqref{DynSysRed}.  This procedure yields 
\begin{align}\label{XYZ_HSn1}
  \begin{split}
&{\mathbf X}=\mathbb{R}-\frac{\mathbb{R}}{2 (3   \beta  \mathbb{A} \mathbb{R}+1)},\\
   &{\mathbf Y}=\frac{8}{\mathbb{R} (6  \beta \mathbb{A}\mathbb{R}+1)}-\frac{4}{\mathbb{R} (3  \beta  \mathbb{A} \mathbb{R}+1)},\\
      &{\mathbf Z}=-\frac{32}{\mathbb{R}^2 (6  \beta   \mathbb{A}
   \mathbb{R}+1)}+\frac{16}{\mathbb{R}^2 (6  \beta   \mathbb{A}
   \mathbb{R}+1)^2}+\frac{16}{\mathbb{R}^2 (3  \beta  \mathbb{A} \mathbb{R}+1)},
\end{split}
\end{align}
and, therefore,
\begin{equation}\label{DynSysHSn1}
  \begin{split}
 &\DerN{\mathbb{R}}=\frac{1}{12\alpha  \beta \mathbb{A}} \left\{ \alpha \mathbb{K}  (1+30  \beta \mathbb{A} \mathbb{R})-(\mathbb{K}+1)(6 \beta \mathbb{A}  \mathbb{R}+1)^3\right. \\
   &~~~~~~~~\left.+\alpha 
   [18 \beta \mathbb{A}  \mathbb{R} (2 \beta \mathbb{A}  \mathbb{R}^2-\mathbb{R} +3)+1]\right\}\\
   &~~~~~~~~+\frac{\Omega  (6  \beta \mathbb{A} \mathbb{R}+1) \left[1-\alpha +12  \beta \mathbb{A} \mathbb{R}(1+3  \beta \mathbb{A} \mathbb{R})\right]}{12 \alpha  \beta \mathbb{A}},\\
 &\DerN{\Omega}=\frac{\Omega}{\alpha -(6  \beta \mathbb{A} \mathbb{R}+1)^2} \left\{(3 \K-2 \mathbb{R}-3 w+2) [\alpha -1-12 \A \beta  \mathbb{R} (3 \A \beta 
   \mathbb{R}+1)]\right.\\
   &~~~~~~~\left.+6 \alpha  \A \beta  \mathbb{R}^2\right\} -\Omega^2\, ,\\
 &\DerN{\K}=2 \K(\K- \mathbb{R}+1),\\
   &\DerN{\mathbb{A} }=-2\mathbb{A} (2+\K-\mathbb{R}).
 \end{split}
\end{equation}

Differently from the system \eqref{DynSysHS}, \eqref{DynSysHSn1} does not present the invariant submanifold $\mathbb{R}=0$, and the submanifold  $\mathbb{A}=0$ is singular.
The properties of the phase space depends on the values of $\alpha$. 

In particular, for $\alpha\neq1$ the phase space contains three fixed points of the type $\mathcal{H}$ and an additional one $\mathcal{H}_\Omega=\{\mathbb{R}=2,\mathbb{K}=0, \Omega=-(4+3w), \mathbb{A}=-\frac{1}{12\beta}\}$. However, all  of these points have $\mathbb{A}<0$ for $\alpha$ and $\beta$ positive and have to be discarded. Therefore, this version of the Hu-Sawicki model presents a finite phase space structure analogous to the one of $f(R)=R+\alpha R^2$.  The difference is that the Hu-Sawicki model dose not necessarily have asymptotic fixed points of the type $\mathcal H$.

For $\alpha =1$, the situation is very different. In this case the function $f$ reduces to 
\begin{equation}
f(R, R_0)= \frac{\beta R_0 R}{1+\beta R_0 R},
\end{equation}
i.e. the Hilbert Einstein term is eliminated. The phase space for this case contains the points  $\mathcal{A}$ (which for $n=1$ coincides with $\mathcal{C}$) to $\mathcal{E}$. Three fixed points of the type $\mathcal{H}$ exist, but two of them  have $\mathbb{A}<0$ for $\beta$ positive and have to be discarded. The third one has coordinates $\{\mathbb{R}=2,\mathbb{K}=0, \Omega=0, \mathbb{A}=0\}$. In addition, two new fixed points with coordinates $\mathcal{I}=\{\mathbb{R}=4,\mathbb{K}=3, \Omega=0, \mathbb{A}=0\}$ and $\mathcal{L}=\{\mathbb{R}=\frac{1}{4}(5-3w),\mathbb{K}=0, \Omega=-\frac{9}{8}(1+w), \mathbb{A}=0\}$ are present. 

The modified Raychaudhuri equation  \eqref{RAy3Ord} returns the same solutions for the fixed points $\mathcal{A}$-$\mathcal{E}$. The solution associated to $\mathcal{I}$ is \eqref{SolABC} whereas the one associated to
$\mathcal{L}$ is 
\begin{equation}
a=a_0(t-t_0)^{\frac{4}{3(1+w)}}.
\end{equation}

\begin{figure}[htbp]
\begin{center}
\includegraphics[scale=0.7]{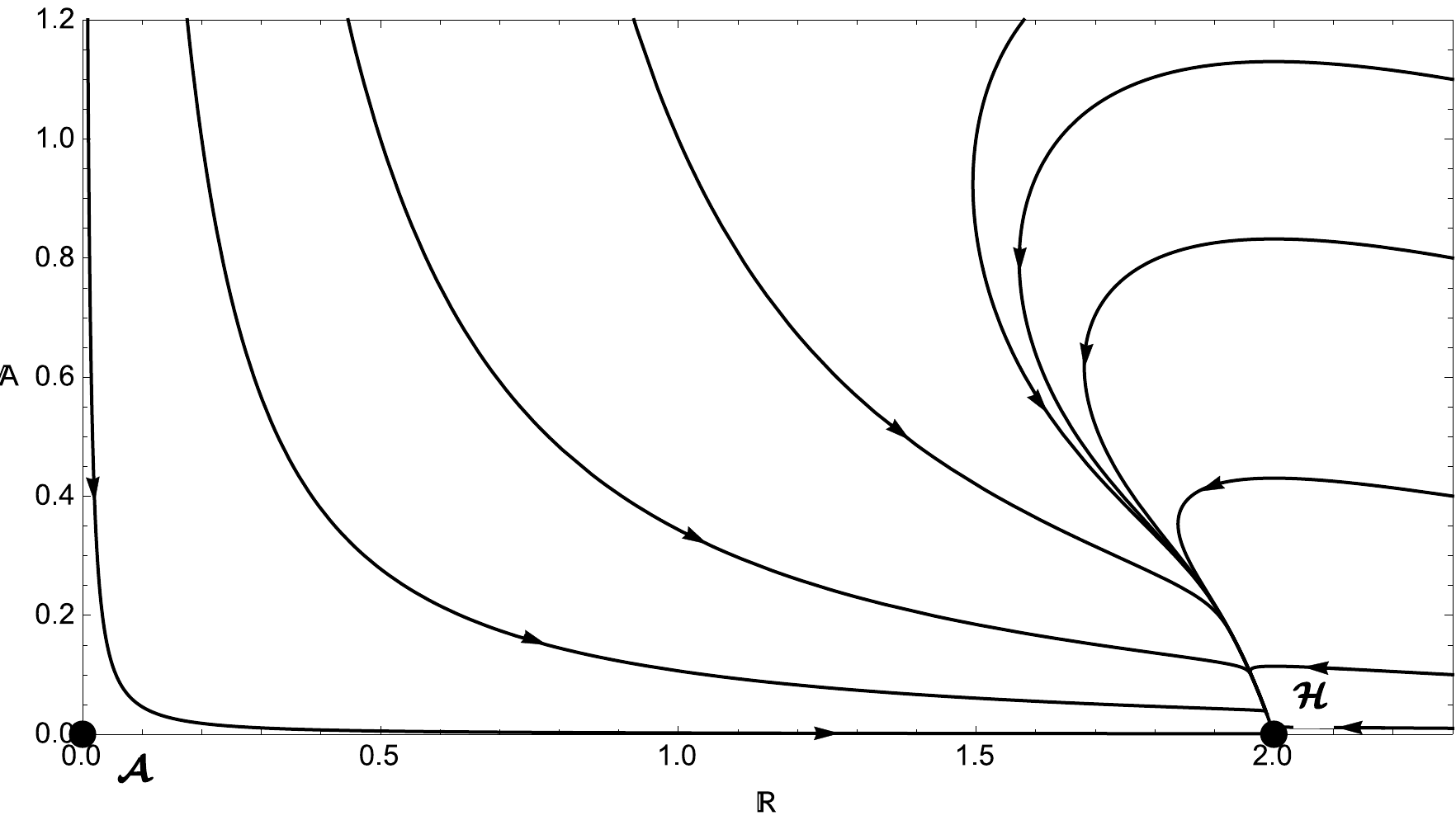}
\caption{The section of the invariant submanifold $\K=0,\Omega=0$  in the case $n=1$, $\alpha=1$, $\beta=\frac{1}{12}$ for the Hu-Sawicki model. Note that on the $\mathbb A>0$ part of this invariant submanifold $\mathcal H$ appears an attractor, but has a saddle character in the unphysical $\mathbb A<0$ of the phase space (not represented here). }
\label{PlotHSn1alp1}
\end{center}
\end{figure}

The stability of the fixed points in this case can be calculated in the standard way and with the exception of $\mathcal{H}$ all appear to be unstable (see Table \ref{HSn1StabAlph1}). Point $\mathcal{H}$ has instead one zero eigenvalue and the analysis of its stability requires the use of the Center Manifold Theorem (CMT) \cite{Car}. To apply this theorem the first step is to write the system \eqref{DynSysHSn1}  in the form
\begin{eqnarray}
&&\DerN{\mathbb A}= C {\mathbb A}  +F({\mathbb A}, {\mathbb Y})\\
&&\DerN{\mathbb Y}= {\mathbf P} {\mathbb Y}+{\mathbf G}({\mathbb A}, {\mathbb Y})
\end{eqnarray}
where  ${\mathbb Y}=\{\mathbb K,\bar{\mathbb R}, \Omega\}$, $\bar{\mathbb R}=3(2-\mathbb R)-12\beta\mathbb A $, $C$  corresponds to the linear pat of the equation for $\mathbb A$, the vector $\mathbf P$  to the linear part of the equation of $\mathbb Y$ and $F$ and  the vector ${\mathbf G}$ represent the non--linear part of the equation for $\mathbb A$ and $\mathbb Y$.  The CMT tells us that the behaviour of  the fixed points is determined by the solution $h$ of the equation
\begin{equation}\label{CMTeq}
\DerN{\mathbb A}= C {\mathbb A}  +F({\mathbb A}, {\mathbf h}({\mathbb A}))
\end{equation}
where the vector  function ${\mathbf h}({\mathbb A})$ is given by 
\begin{equation}\label{eqh1}
{\mathbf h}'({\mathbb A} )\left[C {\mathbb A}   +F({\mathbb A} , h({\mathbb A} )\right]- \left[{\mathbf P} h({\mathbb A} )+{\mathbf G}({\mathbb A} ,h({\mathbb A} ))\right]=0
\end{equation}
Approximating the function ${\mathbf h}({\mathbb A})$ with its Taylor series one finds that 
\begin{equation}
{\mathbf h}({\mathbb A})={\mathbf a} {\mathbb A}^2+...
\end{equation}
where $\mathbf a$ is a vector that depends on the parameter $\beta$. Since the first non-zero term in the previous expression is quadratic, \eqref{CMTeq} implies that point $\mathcal{H}$  has the stability of a saddle-node. In Figure \ref{PlotHSn1alp1} we plot an example of the invariant submanifold $\K=0,\Omega=0$ corresponding to this case.

 \begin{table}[h]
\begin{center}
\caption{Stability of the fixed points of f the Hu-Sawicki model  in the case $n=1$ and $\alpha=1$. Here A stays for attractor,  S for saddle,  F$_A$ for attractive focus and SN for non hyperbolic saddle-node.} \label{HSn1StabAlph1}
 \begin{tabular}{lcccccccccccccc} \hline\hline
Point & Coordinates $\{\mathbb{R},\mathbb{K},\Omega, \A\}$& Scale factor &Stability 
\\ \hline\\
$\mathcal{A}$ & $\left\{ 0, -1 , 0,0\right\}$  & \eqref{SolABC} & S \\ \\
$\mathcal{B}$ & $\left\{ 0, -1 ,-1- 3w,0\right\}$  & \eqref{SolABC}& S   \\ \\
$\mathcal{D}$ & $\left\{ 0, 0 ,2- 3w,0\right\}$  & \eqref{SolDE}& S   \\ \\
$\mathcal{E}$ & $\left\{ 0,0, 0,0\right\}$  & \eqref{SolDE}&  S  \\ \\
$\mathcal{H}$ & $\left\{2,0, 0,0\right\}$  & \eqref{SolDE}&   SN \\ \\
 $\mathcal{I}$ &$\left\{4, 3, 0, 0\right\}$& \eqref{SolH} & S\\ \\
$\mathcal{L}$ &$\left\{\frac{1}{4}(5-3w), 0, -\frac{9}{8}(1+w),0\right\}$& $a=a_0(t-t_0)^{\frac{4}{3(1+w)}}$ & S\\ \\
\hline\hline\\
 \end{tabular}
  \end{center}
\end{table}

\section{Conclusions}
In this paper we have presented a new approach to analyse the finite phase space of the cosmology of $f(R)$-gravity. The new method can be applied to any form of the function $f$ which is analytical in $R$ without the need of cumbersome  inversions. The phase space obtained is more regular than the one obtained with the original dynamical systems approach, although it is not possible to eliminate singularities in a complete way as they are due to the essential form of the cosmological equations.  The new method also naturally excludes the fixed points which represent states incompatible with the definition of the dynamical system variables. Therefore the new DSA returns a phase space which matches in a closer way the actual evolution of the cosmological equations.
  
Using the idea of higher order cosmological parameters (like ``jolt'' and ``snap''), the new DSA is able to  associate to the fixed points full solutions of the cosmological equations (in the sense of solutions with four integration constants).  Among the fixed points found in our analysis, the points labeled $\mathcal H_i$  are surely the most interesting. They  correspond to the dominance of the curvature terms and their number depends on the value of key parameters appearing in the function $f$. The scale factor in $\mathcal H$ is a transcendental  function with the remarkable property to be able to combine an initial exponential expansion a phase of decelerated expansion and a final accelerated expansion phase. Its existence on one hand describes clearly the role of $f(R)$-gravity as model for double inflation or a unified model for inflation, standard cosmology and dark energy. On the other, however, it confirms clearly that $f(R)$ cosmologies with these properties can run into finite time singularities. This is indicated by the fact that in many of the cases we have analysed, the solution $\mathcal H$ is the only attractive fixed point of the phase space (although not a global attractor). It is easy to prove with our method, and in accordance to the results in literature (see e.g. \cite{Capozziello:2009hc,Appleby:2009uf}) that adding a special set of additional invariants one might avoid this scenario, but in general such avoidance requires fine-tuning.

It is important to be careful in considering the nature of the points $\mathcal H$. Their associated solution \eqref{SolH} contains integration constant which can take any value and it is in general related to the constants appearing in $f$. This means that the fact that a theory posses an attractor $\mathcal H$ does not mean that the full behaviour  \eqref{SolH} is necessarily realised: one could have, for example, that for specific values of the coupling constants and of the $H_i$ vanishes. If $H_2$ and $H_3$ are zero then $\mathcal H$ represents a standard de Sitter solution. This depends on the value of the coupling constant of the model as well as the initial condition for the model.  However it is evident that the nature of \eqref{SolH} has repercussions on the understanding of the actual meaning and role of the de Sitter solutions in the context of $f(R)$- gravity.

The DSA proposed has first been verified on the simple case of $f(R)=R^{n}$ and we found a complete agreement in terms of fixed points and stability with the original DSA in \cite{Carloni:2004kp}. However, the solutions associated to the fixed points present significative differences. In fact, even if the solutions obtained with the new method reduce to the ones of the old DSA setting to zero two of the four integration constants, it appears clear that their (time-)asymptotic behaviour is different. This means that, for example, when the fixed points are attractors, the parts of the solution excluded in the original DSA can be dominant. In this respect, therefore, the original  DSA returns incomplete information on the cosmology of $f(R)$-gravity.

As another test of the new DSA, we have considered another form of $f$  that was  analysed with the original dynamical system method: $f(R)=R+\alpha R^n$. In this case, it turns out that the only difference between the two treatment is in the number of fixed points: the new method excludes fixed points that give conditions inconsistent with the definition of the variables. The remaining fixed points, however, present a stability that is completely equivalent.  The theory $f(R)=R+\alpha R^n$ is also the simplest theory that shows points of the type $\mathcal H$ in the finite phase space. An interesting exception in this respect is the case $n=2$ and it is natural to expect that this difference is related somehow to the special properties of this model.

As last steps we have applied the new DSA to two models which were not analysable with the original method, but at the same time constitute important theoretical models  for inflation and/or dark energy: the Starobinsky and the Hu-Sawicki models.  

The Starobinsky model, unfortunately, cannot be treated in general due to the complexity of the algebraic equations needed to determine the fixed points.  We have therefore limited our analysis to the case in which the value of the parameters are chosen to be  compatible with Solar System and cosmological perturbations constraints given in \cite{Starobinsky:2007hu}. In this specific case, we obtain a phase space in which  only  two fixed points of the type $\mathcal H$ appear. Since only one of the points is an attractor we can conclude that the scenario of solution \eqref{solH} is possible in this model, but it is not achievable for general initial conditions. 

The case of the Hu-Sawicki model is more involved. Only for specific intervals of the parameter $n$ the phase space admits fixed points different form the type $\mathcal H$. The situation is complicated by the fact that the points belong to singular submanifold. The case $n=1$ has to be treated separately and reserves a number of surprises. For example, in the case $n=1$, $\alpha\neq 1$ the phase space has no finite fixed points, much in the same way of the case $f(R)=R+\alpha R^2$. The two models, however, differ in the asymptotic structure of the phase space. In the case $n=1$, $\alpha=1$ we found, instead, that only the (unique) point $\mathcal H$ is always effectively an attractor.

The new DSA seems therefore to be very efficient in uncovering the features of interesting $f(R)$ cosmologies. However, as all methodologies, the new DSA presents also a series of drawbacks. For example, one would like to be able to consider in an easier way the GR-like states for the cosmology to be able to find (if they exist) "mimicking behaviours" of these theories, not dissimilar to the isotropization mechanism already found in scalar tensor gravity \cite{isoGR,Carloni:2007eu}. In addition, the impossibility to define a set of compact variables and the fact that our results point clearly to the presence of asymptotic fixed points makes the present analysis incomplete. The resolution of these issues in the context of the new DSA will be the focus of future studies.

\section*{Acknowledgements}
This work was supported by  the Funda\c{c}\~{a}o para a Ci\^{e}ncia e Tecnologia through project IF/00250/2013. The author would like to thank Dr S. Vignolo for useful discussions. 


\end{document}